\pdfoutput=1
\documentclass[aps,prd,amsmath,floats,floatfix, twocolumn,
superscriptaddress,nofootinbib,showpacs,longbibliography]{revtex4-1}

\usepackage[T1]{fontenc}
\usepackage[utf8]{inputenc}
\usepackage{lmodern}
\usepackage{breqn}
\usepackage{bm}
\usepackage{verbatim}

\usepackage[dvipsnames, usenames]{xcolor}
\definecolor{linkcolor}{rgb}{0.0,0.3,0.5}
\usepackage[hypertexnames=false, unicode, colorlinks=true, linkcolor=linkcolor,
citecolor=linkcolor, filecolor=linkcolor,urlcolor=linkcolor,
pdfusetitle]{hyperref}

\usepackage[all]{hypcap}
\usepackage{graphicx}
\usepackage{xspace}
\usepackage{amssymb}
\usepackage{amsmath,amssymb,graphicx}
\usepackage{amssymb,amsmath}

\usepackage{microtype}

\usepackage[english]{babel}
\usepackage{blindtext}

\usepackage[normalem]{ulem} 

\DeclareMathAlphabet{\mathpzc}{OT1}{pzc}{m}{it}

\usepackage{blindtext}

\graphicspath{
	{figs/}
}

\begin{document}
	
	\title{Survey of gravitational wave memory in intermediate mass ratio binaries}
	
	\newcommand{\UMassDMath}{\affiliation{Department of Mathematics,
			University of Massachusetts, Dartmouth, MA 02747, USA}}
	\newcommand{\UMassDPhy}{\affiliation{Department of Physics,
			University of Massachusetts, Dartmouth, MA 02747, USA}}
	\newcommand{\CSCVR}{\affiliation{Center for Scientific Computing and Visualization 	Research, University of Massachusetts, Dartmouth, MA 02747, USA}}
	\newcommand{\URI}{\affiliation{Department of Physics, 
		University of Rhode Island, Kingston, RI 02881, USA}}    
	\newcommand{\UCD}{\affiliation{School  of  Mathematics  and  Statistics,  
			University  College  Dublin,  Belfield,  Dublin  4,  Ireland}}
	\newcommand{\KITP}{\affiliation{Kavli Institute for Theoretical Physics, University of California, Santa Barbara, CA 93106, USA}} 
	
	\author{Tousif Islam}
	\email{tislam@umassd.edu}
	\UMassDPhy
	\UMassDMath
	\CSCVR
	\KITP
	
	\author{Scott E. Field}
	\UMassDMath
	\CSCVR
	
	\author{Gaurav Khanna}
	\UMassDPhy
	\CSCVR
	\URI
	
	\author{Niels Warburton}
	\UCD
		
	\hypersetup{pdfauthor={Islam et al.}}
	
	\date{\today}
	
\begin{abstract}
The non-linear gravitational wave (GW) memory effect is a distinct prediction in general relativity.
While the effect has been well studied for comparable mass binaries, it has mostly been overlooked for intermediate mass ratio inspirals (IMRIs). We offer a comprehensive analysis of the phenomenology and detectability of memory effects, including contributions from subdominant harmonic modes, in heavy IMRIs consisting of a stellar mass black hole and an intermediate mass black hole.
When formed through hierarchical mergers, for example when a GW190521-like remnant captures a stellar mass black hole, IMRI systems have a large total mass, large spin on the primary, and possibly residual eccentricity; features that potentially raise the prospect for memory detection. We compute both the displacement and spin non-linear GW memory from the $m \neq 0$ gravitational waveforms computed within a black hole perturbation theory framework that is partially calibrated to numerical relativity waveforms. We probe the dependence of memory effects on mass ratio, spin, and eccentricity and consider the detectability of a memory signal from IMRIs using current and future GW detectors.
We find that (i) while eccentricity introduces additional features in both displacement and spin memory, it does not appreciatively change the prospects of detectability, (ii) including higher modes into the memory computation can increase signal-to-noise (SNR) values by about 7\% in some cases, (iii) the SNR from displacement memory dramatically increases as the spin approaches large, positive values, (iv) spin memory from heavy IMRIs would, however, be difficult to detect with future generation detectors even from highly spinning systems. Our results suggest that hierarchical binary black hole mergers may be a promising source for detecting memory and could favorably impact memory forecasts.
\end{abstract}

\maketitle

\section{Introduction} 
\label{Sec:Introduction}
Detection of gravitational waves (GWs) from the coalescence of binary compact objects \cite{LIGOScientific:2018mvr,abbott2021gwtc}, mostly binary black-holes (BBH), not only helps to shape our understanding about compact objects in the universe, they also provide an unique opportunity to test the predictions of the general relativity (GR) \cite{Abbott:2020jks}. One important prediction of GR is the existence of gravitational wave memory, a permanent distortion of an idealized GW detector after the wave has passed by \cite{Zeldovich:1974gvh,braginsky1987gravitational,christodoulou1991nonlinear,Blanchet:1992br}. Detection of GW memory would further bolster the validity of GR, while signatures of GW memory in alternative theories of GR are expected to deviate from GR predictions \cite{Kilicarslan:2018bia}. Memory effects, therefore, offer a unique way to probe the  nature of gravity \cite{Yang:2018ceq}. 

GW memory effects could be of different types having distinct origins. Earlier studies have primarily focused on \textit{displacement memory}, which is a lasting change on the gravitational wave strain. \textit{Spin memory} \cite{Pasterski:2015tva} is sourced from the fluxes of the angular momentum of the binary while \textit{center-of-mass memory} \cite{Nichols:2018qac} effect is related to changes in the center-of-mass part of the angular momentum. For typical astrophysical sources, displacement memory is expected to be the dominant effect followed by spin memory and center-of-mass memory \cite{Nichols:2018qac}. 

The displacement memory effect is the most well studied flavor of memory. Calculation of the displacement memory requires the computation of an angular integral of the gravitational-wave energy flux  (see, for example, Eq.~(1) of Ref.~\cite{Talbot:2018sgr}). 
Various approximations and models for this integral have appeared in the literature.
Earlier works \cite{Wiseman:1991ss,Kennefick:1994nw} have used a post-Newtonian quadrupolar approximation to compute displacement memory effects for BBH inspirals. Subsequent studies \cite{Favata:2009ii,Favata:2008yd,Favata:2010zu} extended the memory calculation based on quadrupole modes to the full inspiral-merger-ringdown signal. 

For purely numerical computations of the gravitational wave strain, Refs.~\cite{Johnson:2018xly,Burko:2020gse} used detector-adapted coordinates 
and an approximate Isaacson stress-energy tensor to simplify angular integrals.
The resulting kludge model \cite{Johnson:2018xly,Burko:2020gse} is very simple and applicable to higher harmonic modes, but due to these approximations it is mainly used to study memory phenomenology~\cite{Boersma:2020gxx}. 
For example, this approach has been used to compute the memory effects for zoom-whirl extreme mass ratio inspirals (EMRI) orbits around fast spinning Kerr black holes \cite{Burko:2020gse}. 
Recently Talbot et al.~\cite{Talbot:2018sgr} performed a direct computation of the memory up to $\ell=4$ using full inspiral-merger-ringdown waveforms from numerical relativity simulations including many subdominant modes. 
This direct computation takes into account coupling between modes in the energy flux expression, and it is 
expected to be the most accurate because it introduces the fewest approximations.
We note that NR techniques have largely been unable to extract memory modes \cite{favata2009post},
although recent advances~\cite{Mitman:2020pbt} have made this possible by using
the {\tt SpECTRE} code's~\cite{kidder2017spectre} Cauchy-characteristic extraction module~\cite{moxon2020improved}.

Detectability of memory effects in current and future generation GW detectors has attracted significant interest \cite{favata2009post,Nichols:2017rqr,Lasky:2016knh,McNeill:2017uvq,Johnson:2018xly,Boersma:2020gxx,Hubner:2019sly,Khera:2020mcz,Wang:2014zls}. 
In particular, Ref.~\cite{Johnson:2018xly} reports that third generation detectors will be able to detect memory effects from optimally oriented GW150914-like \cite{Abbott:2016blz} events. 
Ref.~\cite{McNeill:2017uvq} has considered detecting memory effects without being able to detect the ``parent" oscillatory waveform while Ref. \cite{Lasky:2016knh} looks for the evidence of memory in a population of GW150914-like BBH events.
Recently Ref.~\cite{Khera:2020mcz} takes a different approach and attempts to infer a total memory observable from GW events. 
With current generation detectors,
Ref.~\cite{Hubner:2019sly} finds that $\sim 2000$ GW events need to be combined in order to recover strong evidence for memory effects in a binary population while Ref.~\cite{Boersma:2020gxx} estimates that it would take $\sim5$ years of data for the memory modes to reach an SNR threshold of 3 in current detectors. Consistent with these expectations, no evidence of memory has been found in the population of the 50 GW BBH merger events reported in the first and second LIGO/Virgo gravitational-wave transient catalogs~\cite{hubner2021memory}.

Previous studies have mostly focused on BBH systems that are representative of the events found in the first and second observing runs~\cite{LIGOScientific:2018mvr}. Similar to the first detection, GW150914, these are comparable mass, moderately (or non-) spinning systems in quasi-circular orbit.
Here we consider heavy intermediate mass ratio inspirals (IMRIs) with the possibility of high spin and eccentricity.
These binaries consist of an intermediate mass black hole (IMBH) with mass $\sim10^2-10^4 M_\odot$ \cite{Feng:2011pc,Pasham:2015tca,Mezcua:2017npy} paired with a first-generation black hole produced by stellar collapse of mass $\sim4-40M_\odot$.  The resulting binaries will have a mass ratio in the range $q:=m_1/m_2= 2-10^4:1$ where $m_{1}$ ($m_2$) is the mass of the more (less) massive black hole. Evidence for IMBHs was previously known through indirect electromagnetic observations \cite{2017IJMPD..2630021M} and the low-end of this mass range has recently been confirmed by GW190521 \cite{Abbott:2020tfl}. The GW190521 remnant black hole has an estimated mass and dimensionless spin of $142^{+28}_{-16} M_\odot$ and $0.72^{+0.09}_{-0.12}$, respectively. 

When formed through hierarchical mergers~\cite{doctor2020black,gerosa2021hierarchical}, for example when a GW190521-like remnant captures a stellar mass black hole, IMRI systems typically have a large total mass, large spin on the primary, and possibly residual eccentricity; features that potentially raise the prospect for memory detection especially when subdominant modes are included into the analysis \cite{Talbot:2018sgr,Liu:2021zys}. On the other hand, a competing effect is that higher-mass-ratio sources emit weaker signals. A key goal of this paper is to explore the dependence of the memory's SNR as the total mass, spin, mass ratio, and eccentricity parameters are varied for different IMRI configurations similar to what might be expected if a GW190521-like remnant captured a stellar-mass BH. IMRIs are likely to exist in dense globular clusters and galactic nuclei \cite{Leigh:2014oda,MacLeod:2015bpa} and are one of the prime sources for future-generation detectors \cite{AmaroSeoane:2007aw,Berry:2019wgg}. In particular, IMRIs with total mass $<2000M_\odot$ may be detected by current-generation detectors \cite{Amaro-Seoane:2018gbb} with higher mass binaries detectable by future space-based missions such as LISA \cite{2017arXiv170200786A} and beyond \cite{Kawamura:2020pcg, Sedda:2019uro}. 

The rest of the paper is organized as follows. In Sec.~\ref{sec:nonlinear_GW_memory}, we provide an overview of the models used to compute the memory effects. Sec.~\ref{imridata} describes our method for computing IMRI waveforms. In Sec.~\ref{robustness_study}, we assess potential sources of systematic error in our memory computation. We then present results for the memory signal and its dependence on spin, eccentricity, and mass ratio (Sec.~\ref{phenomenology}) as well as the signal-to-noise ratio across different detectors. Finally, we discuss the implication of our results and caveats in Sec.~\ref{Sec:Conclusion}.

\section{Nonlinear gravitational wave memory models} 
\label{sec:nonlinear_GW_memory}
We model both the displacement and spin memory components. While displacement memory contributes predominantly to the $h^{\rm mem}_{20}$ and $h^{\rm  mem}_{40}$ spherical harmonics modes (and therefore to plus polarization of the memory waveform), spin memory contributions show up in the $h^{\rm mem}_{30}$ mode (and therefore in the cross polarization) \cite{Nichols:2017rqr}. Below we describe the computation of both memory effects from a given oscillatory gravitational waveform.

\subsection{Computing Displacement Memory}
\label{dmm}
The non-linear displacement memory sourced by  gravitational waves can be computed using the expression~\cite{Wiseman:1991ss,Thorne:1992sdb,Favata:2010zu}
\begin{equation}\label{eq:memory}
	h^{TT,\rm dis}_{jk}(T_r, r, \Omega)=\frac{4}{r}\int_{-\infty}^{T_r}dt \int_{S^{2}} d\Omega^{\prime} \frac{dE}{dt d\Omega^{\prime}} \left[ \frac{n_j^{\prime} n_k^{\prime}}{1-n_l^{\prime}n^{l}} \right]^{TT},
\end{equation}
where $r$ is the distance between the source and the observer, $T_{r}$ is the retarded time, $\Omega=(\iota, \phi)$ are the angles ($\iota$ is the inclination angle between the angular momentum vector of the binary and the line of sight of the observer, $\phi$ is the reference phase at coalescence), and $n(\Omega)$ is the unit vector pointing from
the source to the observer located at $\Omega$. Here, $\Omega^{\prime}$ coordinates describe the sphere over which the integral is taken and $n'(\Omega')$ is the associated unit vector.

The gravitational wave energy flux, $\frac{dE}{dt d\Omega}$, can be written in terms of the time-derivatives of the GW strain,
\begin{equation}\label{eq:fluxharmonic}
	\frac{d E}{dt d\Omega} = \frac{r^2}{16\pi}\sum\limits_{\ell',\ell'',m',m''} \Big \langle \dot{h}_{\ell' m'}\dot{h}^{*}_{\ell''m''}  \Big \rangle \ {^{(-2)}Y_{\ell' m'}}\  ^{(-2)}Y_{\ell''m''}^{*},
\end{equation}
expanded in spin-weighted spherical harmonics $^{(-2)}Y_{\ell m}$, where $\langle.\rangle$ denotes an average over a few waveform cycles\footnote{In practice, such waveform averaging has not been done in any of the models considered in this paper as the memory is known to be relatively insensitive to this procedure~\cite{Favata:2011qi}.} and $h^{*}_{\ell m}$ indicates the complex conjugate of $h_{\ell m}$.
The memory expression in Eq.~\eqref{eq:memory} is then projected onto the two orthogonal polarizations of the GWs by contracting with the complex polarization tensors $e_{+}^{ij}$, $e_{\times}^{ij}$:
\begin{equation} \label{eq:hmemPols}
	h^{\rm dis}  = h_{+}^{\rm dis} - i h_{\times}^{\rm dis} = h_{jk}^{TT,\rm dis} (e_+^{jk} - i e_\times^{jk}) \, .
\end{equation}
It is then convenient to expand the displacement memory
\begin{equation} \label{eq:hmempx}
	h^{\rm dis}  = \sum_{\ell,m} h_{\ell m}^{\rm dis} \ ^{(-2)}Y_{\ell m} \, ,
\end{equation}
in terms of spin-weighted  spherical harmonics. 
Below we summarize three approximate methods used in the literature to compute Eq.~\eqref{eq:hmempx}.
\subsubsection{Quadrupole approximation}
This model uses the dominant (oscillatory) quadrupole mode $h_{22}$ to compute the displacement memory contributions~\cite{Favata:2010zu},
\begin{eqnarray}\label{dmm_model}
	h^{\rm{(dis)}}_{+\;2 2}(t)=\frac{1}{192\pi r}\Phi_1(\iota)\int_{-\infty}^{t}\,dt' \left|\dot{h}_{2 2}(t')\right|^2\, ,
\end{eqnarray}
where $\Phi_1(\iota):=\,\sin^2\iota\,(17+\,\cos^2\iota)$, an overdot denotes differentiation with respect to $t'$, and $|\cdot|$ denotes the complex modulus. For displacement memory sourced by the oscillatory (2,2) mode, $h^{\rm{(dis)}}_{\times \;2 2} = 0$. The quadrupole model has often been used to study the phenomenology of displacement memory \cite{Favata:2009ii,Johnson:2018xly,Burko:2020gse}.
\subsubsection{Minimal waveform  model}
The minimal waveform model ($mwm$) \cite{Favata:2009ii} employs an analytical expression combining a PN approximation for the inspiral and a superposition of quasinormal modes during the merger and ringdown to compute the displacement memory both in time and frequency domain. 
The model has been calibrated to an effective-one-body (EOB) model to fit the nonlinear displacement memory in the comparable mass ratio regime.
It is known that the $mwm$ model over estimates the memory contribution \cite{Talbot:2018sgr,Boersma:2020gxx,Favata:2009ii}.
Nonetheless, for the sake of completeness, in Sec.~\ref{sec:mem_models} we include this model in our comparison study.

\subsubsection{Higher multipole  model}
The \textit{higher multipole model}, derived by Talbot et al.~\cite{Talbot:2018sgr}, evaluates the expression in Eq.~\eqref{eq:hmempx} by numerically integrating Eq.~\eqref{eq:memory}. 
The model uses both the quadrupole mode as well as available higher order modes up to $\ell=4$ and accounts for the coupling between modes in the energy flux expression. 
The python package $\texttt{GWMemory}$ \cite{gwmemory} used to carry out the computation is publicly available~\footnote{The specific version of the code we use has a commit hash of 2a4b8084144b13a3f542b1e132d9bc629d4ec9c1. In particular, pull requests 7 and 19 correct various data files used in the memory computation. The paper's first version on the arXiv used a version of the code before pull request 7.}. 
Unless otherwise mentioned, we use this higher multiple model from $\texttt{GWMemory}$
for calculating displacement memory. In Appendix~\ref{app:code-comparison} we compare $\texttt{GWMemory}$ to two additional higher multipole memory models, finding that all three perform similarly.

\subsection{Computing Spin Memory}
\label{smm}
Following \cite{Nichols:2017rqr}, the spin memory
\begin{eqnarray}\label{smm_model}
	h^{\rm{(spin)}}_{\times\;2 2}(t)=\frac{3}{64\pi r}\Phi_2(\iota) \mathcal{I}( {U}^{*}_{22}\dot{U}_{22}).
\end{eqnarray}
is computed directly from the oscillatory (2,2) mode of the GW signal. Here, $\Phi_2(\iota):=\,\sin^2\iota\cos\iota$, $\mathcal{I}$ denotes the imaginary part, and $U_{22}$ is given by:
\begin{eqnarray}\label{Ulm}
U_{22}(t)=\frac{1}{\sqrt{2}}[h_{22}+h^{*}_{2,-2}] \,.
\end{eqnarray}

\section{IMRI waveform model}
\label{imridata}
Modeling the late inspiral and merger regimes from IMRI systems is challenging. 
One reason is that these systems are essentially inaccessible to exploration by 
numerical relativity codes due to the small length scale introduced by the lighter black hole. 
This is because the inspiral time scales linearly with $q$ and finer grid resolution is required to resolve a smaller secondary. For these reasons the majority of NR simulations have had $q\le 15$ \cite{Boyle:2019kee, Healy:2020vre} with a small handful of short-duration simulations performed at higher mass ratios \cite{Lousto:2020tnb}. 
As modern gravitational-waveform modeling efforts require NR data for calibration, 
a lack of NR data in this regime has prevented the construction of extensive and accurate models.
One potential path forward was recently developed and
applied to nonspinning, quasi-circular BBH systems~\cite{Rifat:2019ltp}. 
We now summarize our simple application of this technique 
to spinning and eccentric IMRI systems.

We generate our gravitational waveforms using black hole perturbation theory (BHPT). In this approach,  
the smaller black hole with a mass $m_2$ is modeled as a point particle, with no internal structure, moving in the space-time of the heavier Kerr black hole with mass $m_1$ and spin angular momentum per unit mass $a$. 
The inspiralling trajectory of the particle is computed using standard energy and angular momentum balance equations \cite{10.1143/PTP.112.415,10.1143/PTP.113.1165,10.1143/PTP.95.1079,ThorweThesis}. 
To compute flux radiated to future null infinity and through the event horizon for the quasi-circular inspirals we use the \texttt{Gremlin} code \cite{gremlin,osullivan2014strongfield,Drasco_2006} from the Black Hole Perturbation Toolkit \cite{BHPToolkit}. 
For eccentric orbits we use Schwarzschild flux data which is available at Ref. \cite{BHPToolkit}, and we arrange the inspirals such that they fully circularize before the onset of the plunge. 
The inspiral trajectory is then extended to include the plunging trajectory \cite{ori2000transition,hughes2019learning,sundararajan2010binary,apte2019exciting}. 
With the complete trajectory in hand, the gravitational radiation is computed by numerically solving the time-domain Teukolsky equation in compactified hyperboloidal coordinates \cite{Sundararajan:2007jg,Sundararajan:2008zm,Sundararajan:2010sr,Zenginoglu:2011zz}. The resulting waveforms include the inspiral, merger, and ringdown of the binary. For an executive summary of the methods, see Sec.~III of Ref.~\cite{Burko:2020gse}.

Waveforms computed as sketched above are suitable models for extreme mass-ratio inspirals
where $q \gg 1$. 
Over the past decade, however, there has been mounting evidence that domain of validity of BHPT can be extended to include moderate mass-ratio binaries \cite{LeTiec:2011bk,vandeMeent:2020xgc,Warburton:2021kwk,Wardell:2021fyy}. Recently it was shown waveforms from non-spinning, quasi-circular binaries generated via BHPT can be made to agree remarkably well with NR waveforms with $q\sim3$ --  $15$ via a simple change of the system's mass scale \cite{Rifat:2019ltp}. In order to model IMRIs we follow Ref.~\cite{Rifat:2019ltp} and rescale the BHPT waveforms such that
\begin{equation}\label{alpha}
	h^{\ell m}(t;q, e_{\rm ref}, \chi)=\alpha h^{\ell m}_{\rm BHPT}(t\alpha;q, e_{\rm ref}, \chi),
\end{equation}
where $h^{\ell m}_{BHPT}$ are the spin-weight $-2$ spherical harmonic modes of the waveform
computed from the Teukolsky solver. Here, 
$\chi = a / m_1$ is the dimensionless spin parameter of the heavier black hole, and
$e_{\rm ref}$ is the eccentricity described in Sec.~\ref{sec:eccentricity}. We use $(\ell,m)={(2,2),(2,1),(3,3),(3,2),(3,1),(4,2),(4,3)}$ modes in our computation. Negative $m$ modes are computed using orbital plane symmetry, $h^{\ell,-m}=(-1)^l (h^{\ell m})^{*}$. The contribution of omitted higher order modes are negligible and these modes are often dominated by numerical error, and so we exclude them in this analysis.

The rescaling coefficient
\begin{dmath}
	\alpha(\nu)=1-1.352854\nu - 1.2230006 \nu^2 + 8.601968\nu^3-46.74562\nu^4 \,,
\end{dmath}
used in Eq.~\eqref{alpha} was obtained by fitting the $(2,2)$-mode BHPT waveforms against the $(2,2)$-mode NR data with non-spinning, quasi-circular binaries from mass ratio $q=3$ to $q=10$, where $\nu=q/(1+q)^2$ is the symmetric mass ratio of the binary. 
The rescaled waveform was also shown to agree with a $q=15$, non-spinning NR waveform not used in the fit with a mismatch value of $0.01$ \cite{Rifat:2019ltp}.
As the mass-ratio increases, the scaling factor $\alpha$ approaches unity thereby recovering the fiducial BHPT waveforms.
The calibrated-BHPT waveform approach provides a method for computing IMRI waveforms in a regime currently 
inaccessible to NR simulations.

Thus far the $\alpha$ rescaling has only been determined for non-spinning, quasi-circular binaries. 
Nonetheless we will use it to rescale the low eccentricity and spinning BHPT waveform data we use in this work.
While we do not expect high-accuracy waveforms to be produced by this simple method, for our purpose it will be sufficient for surveying gravitational-wave memory from IMRIs as we demonstrate in the Sec.~\ref{robustness_study}.

\section{Robustness study}
\label{robustness_study}

In this section we explore the robustness of our model for memory from IMRIs. 
In doing so, and for later sections, it will be useful to define the signal to noise ratio (SNR), $\rho$, via
\begin{eqnarray}\label{snr}
	\rho^2=4 \int_{f_{\rm min}}^{f_{\rm max}} \frac{| \tilde{h}(f)|^2}{S_{n}(f)}df,
\end{eqnarray}
where $S_{n}(f)$ is the one-sided noise power-spectral density of the detector, $\tilde{h}(f)$ is the Fourier transform of the detector response given by
\begin{equation}
	h(t)=F_{+}h_{+}+F_{\times}h_{\times} \,,
\end{equation}
and where $F_{+}$ and $F_{\times}$ are the antenna response functions of the detector. 
Our SNR computations will always use a single-detector configuration.
The minimum and maximum frequencies, $f_{\rm min}$ and $f_{\rm max}$, in the limits of Eq.~\eqref{snr} are chosen to reflect the sensitivity bandwidth of the detector. For the detector configurations considered in this paper, we integrate over the frequencies 20 Hz to 1 kHz (for aLIGO and KAGRA \cite{Somiya:2011np}; 20 Hz is the default value used by LIGO and Virgo during the two most recent observing runs~\cite{abbott2021gwtc,abbott2021gwtc:3}), 5 Hz to 1 kHz (for Einstein Telescope \cite{Hild:2010id,maggiore2020science}), and 10 Hz to 1 kHz (for Cosmic Explorer \cite{Evans:2016mbw,hall2021gravitational,hall2021gravitational2})). The lower frequency cutoff for ET and CE are subject to engineering uncertainties, although target design sensitivity values are consistent with ours. In future studies, it would be interesting to explore the impact of achievable sensitivity in the lower frequency band on the memory's SNR.
Before transforming the time domain waveform to the frequency domain, we taper the time domain oscillatory waveform using a Planck window \cite{McKechan:2010kp} while no tapering is used for the memory signal as it introduces additional non-physical features.\footnote{Tapering the time-domain memory waveform introduces Dirac delta function like structure in the merger-ringdown part. This results in a flat plateau region in the frequency domain, which is not a physical feature.}

\begin{figure}[t]
	\includegraphics[width=\columnwidth]{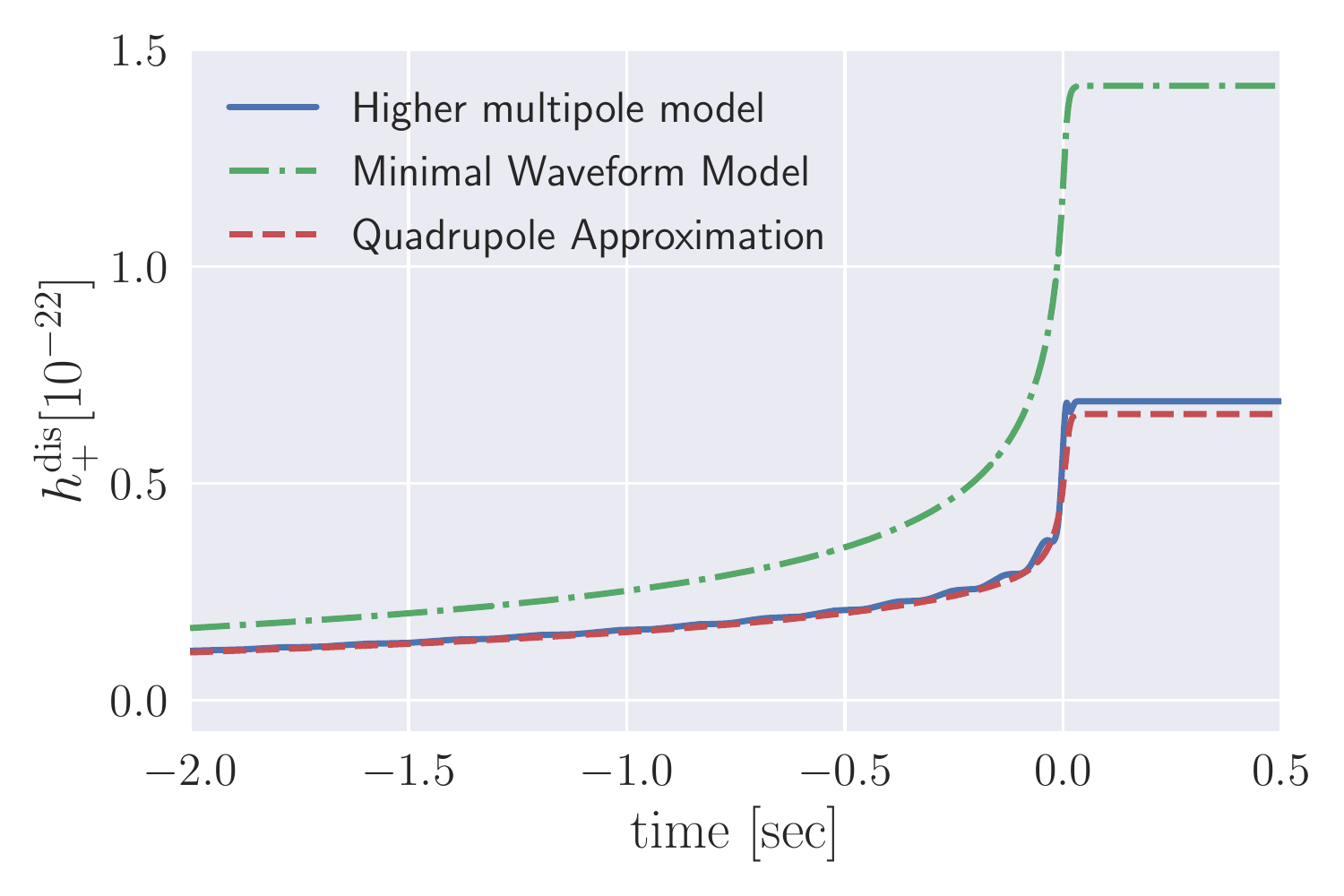}
	\caption{Displacement memory $h^{\rm dis}(t)$ for a non-spinning BBH with total mass $M=200$ $M_{\odot}$, mass ratio $q=10$, a luminosity distance $D=250$Mpc, and at an inclination $\iota=\pi/4$ computed using the three models described in Sec.~\ref{dmm}. Unless otherwise specified, we use detector-frame masses throughout this paper. As has been previously noted~\cite{Boersma:2020gxx}, the \textit{mwm} (dash-dot green line) overestimates the displacement memory effect. The memory signal computed from the dominant (2,2) mode (dashed red line) slightly underestimates the effect as compared to a computation using all available modes (solid blue line). The results of our paper use the higher multipole model~\cite{Talbot:2018sgr,gwmemory} as it is expected to be the most accurate. 
	\label{Fig:MWM}
	}
\end{figure}

\subsection{Effect of truncating the weak field inspiral}
\label{sec:truncating}

Memory effects accumulate over the entire evolution of the binary and formally the lower limit of integration in Eq.~\eqref{eq:memory} is negative infinity.
However, we start the integration at the start of the waveform $t=-10,000M$, where $t=0$ denotes the time at peak waveform amplitude and $M=m_1+m_2$ is the total mass of the binary.
We set the memory signal to zero at the start time.
To obtain a physical strain, the dimensionless waveform is then appropriately scaled using total mass and distance. Unless otherwise specified, this prescription is applied throughout the paper.

In principle, one can use a PN correction to set a non-zero value for $h^{\rm dis}$ at $t=-10,000M$.
We investigate whether such corrections are important to understand the detectability of memory signals.
We generate displacement memory waveforms using the \textit{mwm} model for a binary with $q=10$, $M=100$ $M_{\odot}$, $D=250$ Mpc and $\iota=\pi/4$. Throughout this paper, we use detector frame masses. In one case, we set the initial value of $h^{\rm dis}_{22}(t)$ to zero.
In the other case, we allow the \textit{mwm} waveforms to retain their non-zero values informed by PN terms. The waveforms yield an angle averaged SNR of 0.91 and 1.1 respectively in advanced LIGO detector indicating marginal differences in SNR. We therefore do not use any PN correction in our memory model. 

\subsection{Comparison between different displacement memory approximations} \label{sec:mem_models}
As a first look at our memory calculation, in Fig.~\ref{Fig:MWM} we plot the displacement memory computed using the three models described in Sec.~\ref{sec:nonlinear_GW_memory}.
For a fair comparison, we set $h^{\rm dis}_{22}(t)$ to be zero at the start of the waveform as described in Sec~\ref{sec:truncating}. We find that the time evolution of the memory waveforms are similar for all the models. 
However, memory computed from the quadrupolar mode generated through point-particle black hole perturbation theory exhibits slightly smaller values compared the higher multipole model. 
The \textit{mwm} model, on the other hand, overestimates the memory effects. 
This is not unexpected as the \textit{mwm} model is calibrated for binaries with equal masses and/or small spins.
Similar results were found in, e.g., Ref.~\cite{Boersma:2020gxx} -- see their Fig.~1.

\subsection{Comparison between displacement memory effects computed using ppBHPT and NR waveforms in the comparable mass ratio regime}
\label{nrhyb_comparison}

\begin{figure}[t]
	\includegraphics[scale=0.57]{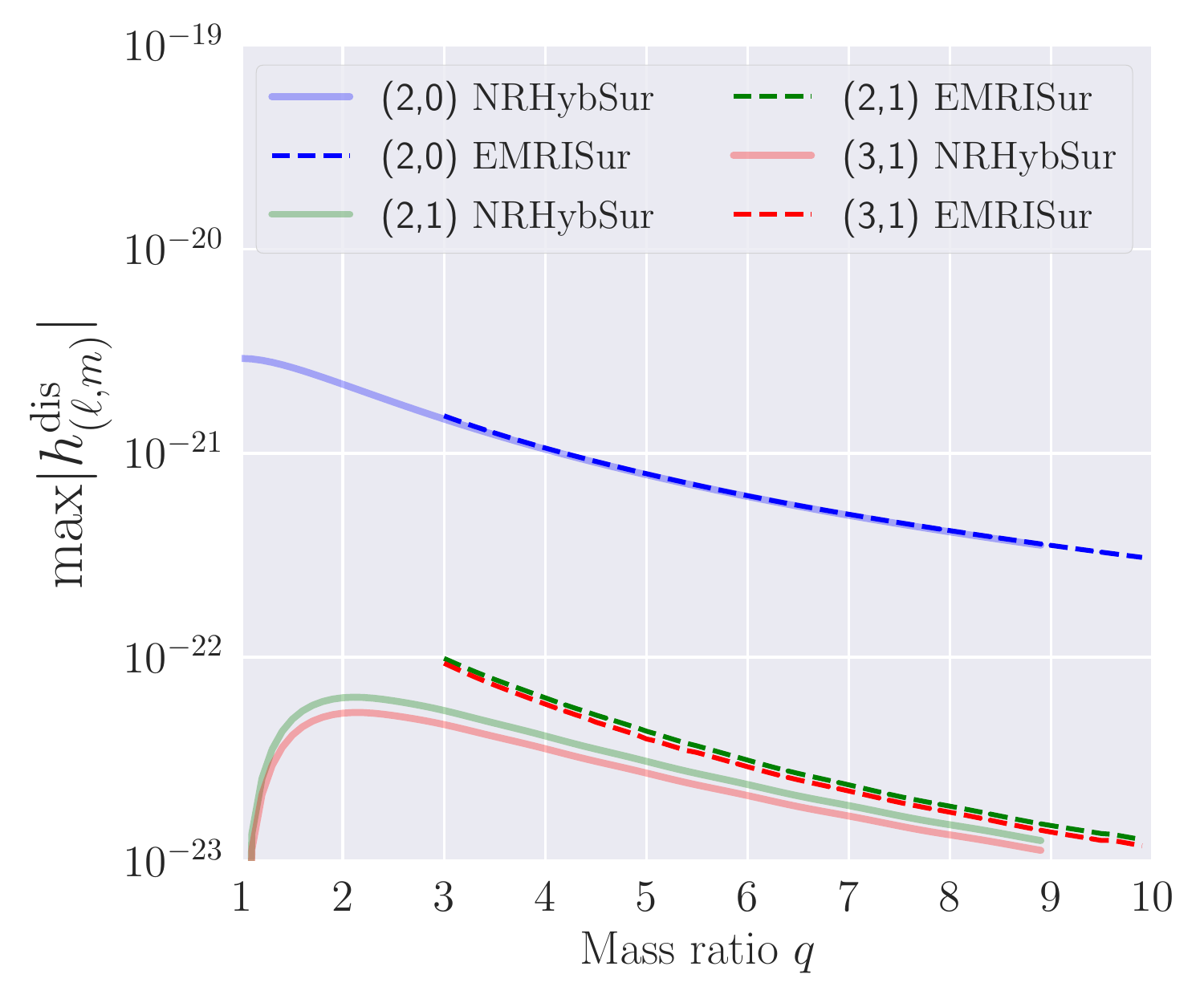}
	\caption{Maximum amplitude of three different memory modes as a function of mass ratio. We compute the displacement memory using two different waveform models: EOB-NR hybridized aligned spin surrogate waveform \texttt{NRHybSur3dq8} (\textit{solid lines; labeled as} \texttt{NRHybSur}) and an $\alpha$-calibrated ppBHPT waveform from the surrogate model \texttt{EMRISur1dq1e4}  (\textit{dashed lines; labeled as} \texttt{EMRISur}). We observe 
	a reasonable match between the memory effects computed with these two different models (details in text; Sec. \ref{nrhyb_comparison}).
	}
	\label{Fig:memory_modes_NRHybSur3dq8}
\end{figure}

As a second check on our memory calculation, we compute the memory modes for different binaries in the small mass ratio regime ($1\le q \le 10$) while fixing $M=200$ $M_{\odot}$, $\chi=0.0$, $e_{\rm ref}=0.0$ and $D=250$ Mpc. We perform this analysis with two different waveform families: a hybridized EOB-NR based aligned-spin surrogate model \texttt{NRHybSur3dq8} \cite{Varma:2018mmi} and \texttt{EMRISur1dq1e4} \cite{Rifat:2019ltp}, a surrogate version of the ppBHPT waveforms calibrated to NR.

Fig.~\ref{Fig:memory_modes_NRHybSur3dq8} shows the dominant $(2,0)$ memory mode along with two important subdominant modes.
For \texttt{NRHybSur3dq8}, we compute the memory modes for $1\le q\le9$ and, for \texttt{EMRISur1dq1e4}, we show the results for $q\ge3$ to reflect their respective domains of validity. 
The figure shows visual consistency in the computation of $(2,0)$ memory mode, but noticeable discrepancies for these subdominant modes.
Differences in the higher order memory modes arise as the higher order oscillatory modes in the $\alpha$-calibrated ppBHPT waveforms are not individually tuned to NR. Similarly, small (but noticeable) differences in the (2,0) mode are due to mode coupling between the subdominant oscillatory modes that arise in the evaluation of Eq.~\eqref{eq:fluxharmonic}; Sec.~\ref{sec:vsEOB} considers a memory computation using only the quadrupole oscillatory mode where we no longer find any discrepancies. 

The small differences shown in Fig.~\ref{Fig:memory_modes_NRHybSur3dq8} are not a concern for our SNR computations as the $(2,0)$ memory mode is expected to be dominant over other
higher order memory modes (see Figs.~\ref{Fig:memory_modes_q10} and~\ref{Fig:memory_modes_q1075}). 
We also observe that the differences in the higher order modes decrease as mass ratio increases as the ppBHPT framework is expected to perform better in the high mass ratio regime that we are interested in. 
Interestingly, we observe that the higher order memory modes have a maxima around $q\sim2$. 
This is consistent with the findings of Talbot et al.~\cite{Talbot:2018sgr} who observe a growth in the maximum of the (2,1) and (3,1) memory modes as q increases from 1 to 2 (cf. Fig.~3 of Ref.~\cite{Talbot:2018sgr}). While we have not explored the origin of this behavior, we note that for nonspinning BBH systems in quasicircular orbit, the odd-m oscillatory modes' amplitude are zero at q=1, turn on quickly as $q$ becomes non-unity, and then transition to $\propto 1/q$ behavior as $q$ becomes large.

Further calibration of the higher order radiative modes in the \texttt{EMRISur1dq1e4}  model \cite{nextgenemri} will improve the agreement between the two models for the higher order memory modes.
The comparison in this section for $q\le10$ shows a good agreement for the dominant $(2,0)$ memory mode and reasonable qualitative agreement for the subdominant memory modes.
As both models are calibrated to NR simulations for $q\le10$, we next consider mass ratios $q \ge 10$.

\subsection{Comparison between displacement memory effects computed using $\alpha$-calibrated ppBHPT and EOB waveforms in intermediate mass ratio regime}
\label{sec:vsEOB}
\begin{figure}[h!]
	\includegraphics[scale=0.5]{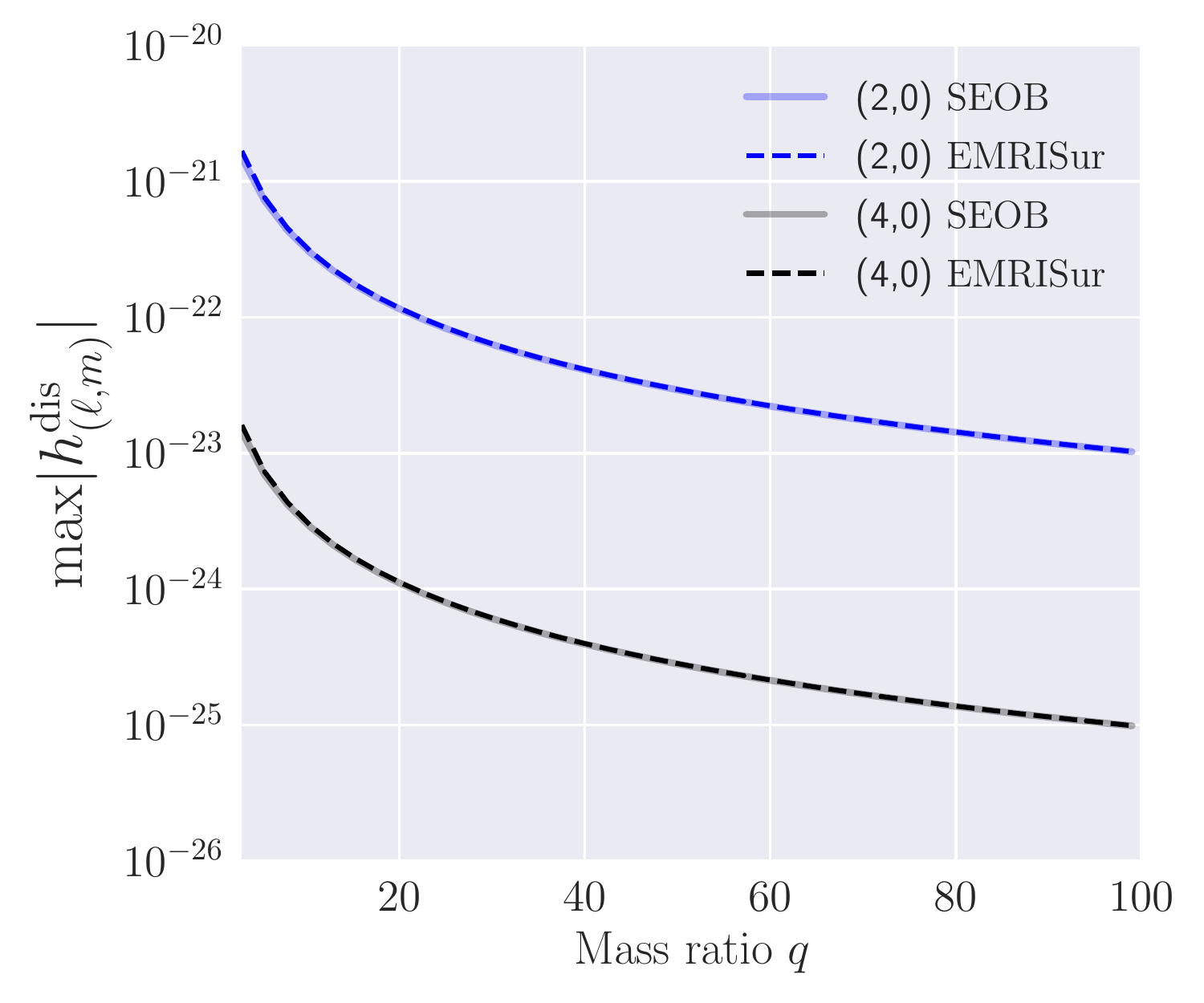}
	\caption{Maximum amplitude of two different memory modes as a function of mass ratio. We compute the displacement memory using two different waveform models: an aligned-spin EOB model \texttt{SEOBNRv4HM} (\textit{solid lines; labeled as} \texttt{SEOB}) and a ppBHPT-based surrogate waveform \texttt{EMRISur1dq1e4}  (\textit{dashed lines; labeled as} \texttt{EMRISur}). We fix $M=200$ $M_{\odot}$, $\chi=0.0$, $e_{\rm ref}=0.0$ and $D=250$ Mpc.
	We find the relative difference for the maximum displacement memory computed using these two models is always less than 1\% for both the $(2,0)$ and $(4,0)$ memory modes.
	The observed agreement between these two different models in both small and intermediate mass ratio regime gives us confidence that the memory effects computed for IMRIs using either \texttt{EMRISur1dq1e4}, \texttt{SEOBNRv4HM}, or $\alpha$-calibrated \texttt{ppBHPT} waveforms accurately capture the true displacement memory in this regime.}
	\label{Fig:memory_modes_SEOB}
\end{figure}

As discussed in Sec.~\ref{imridata} there are essentially no NR simulations of IMRIs.
In lieu of direct comparison to NR, to explore the robustness of our memory calculation for IMRIs we now compute displacement memory for mass ratios $3\le q \le 100$ using the $\alpha$-calibrated ppBHPT waveforms and an aligned-spin EOB model \texttt{SEOBNRv4HM} \cite{Cotesta:2018fcv}.
By construction both models give the correct result in the geodesic, $q\rightarrow\infty$, limit, both include some information from linear-in-the-mass-ratio BHPT, and both are calibrated against NR simulations for $q\le10$.
Neither model is calibrated in the IMRI regime but we find the two models provide similar memory computations from binaries with mass ratios $q\le 100$.
To demonstrate this we use the dominant quadrupolar mode to calculate the memory effects. 
In Fig.~\ref{Fig:memory_modes_SEOB}, we show the maximum amplitude of the displacement memory as a function of mass ratio $q$ where we fix $M=200$ $M_{\odot}$, $\chi=0.0$, $e_{\rm ref}=0.0$, and $D=250$ Mpc. 
Up to $q=100$ we find the relative difference for the maximum displacement memory computed using the two models is always less than 3\% for the dominant $(2,0)$ memory mode.

\begin{figure}[h!]
	\includegraphics[scale=0.5]{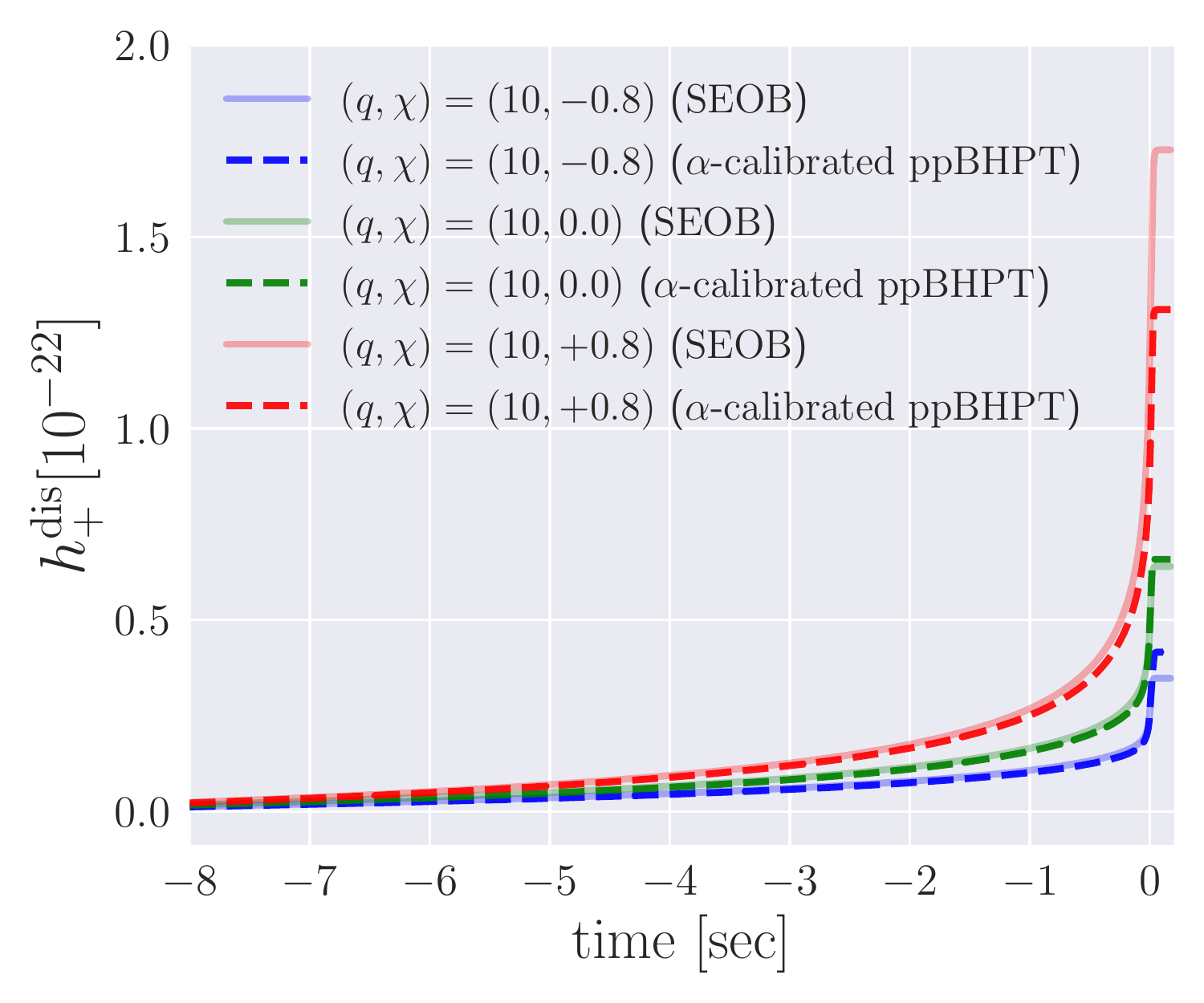}
	\caption{Displacement memory for three different spin configurations  $\chi=\{-0.8,0.0,+0.8\}$ with mass ratio $q=100$, $M=200$ $M_{\odot}$, $\chi=0.0$, $e_{\rm ref}=0.0$ and $D=250$ Mpc. We compute the displacement memory using two different waveform models: an aligned-spin EOB model \texttt{SEOBNRv4HM} (\textit{solid lines; labeled as} \texttt{SEOB}) and a $\alpha$-calibrated ppBHPT waveforms (\textit{dashed lines; labeled as} \texttt{$\alpha$-calibrated ppBHPT)}.
	For $\chi=+0.8$, final displacement memory value between SEOB and \texttt{$\alpha$-calibrated ppBHPT)} are a bit different. Otherwise, relative differences between the displacement memory profiles from these two models are always less than 8\%.
	The observed agreement between the memory effects computed with these two different models gives us confidence that the memory effects computed for IMRIs using \texttt{SEOBNRv4HM} or the $\alpha$-calibrated \texttt{ppBHPT} waveforms accurately capture the true displacement memory in this regime.
	}
	\label{Fig:q10_SEOB}
\end{figure}

We now consider the consistency between the two models when the larger black hole is spinning. 
In this portion of the parameter space the SEOBNRv4HM model has been calibrated using spinning NR simulations.
On the other hand the $\alpha$-calibrated ppBHPT model computes the GWs from an inspiral into a Kerr black hole in the extreme mass ratio limit and then rescales to the waveform for IMRIs using the parameter $\alpha$ which is fitted to NR data for \textit{non-spinning} binaries.
Nonetheless we find that the memory computed for spinning binaries using the $\alpha$-calibrated ppBHPT and SEOBNRv4HM models continue to agree well for spinning binaries.
In Fig.~\ref{Fig:q10_SEOB} we show the total displacement memory computed using the two models for three different spin configurations $\chi=\{-0.8,0.0,+0.8\}$ for mass ratio $q=10$. 
All other parameters have remained the same as those reported in the previous paragraph. 

The observed agreement between these two models gives us confidence in the memory effects computed from both in the IMRI regime.
It is worth noting that while we find that the memory computed using the $\alpha$-calibrated ppBHPT and SEOBNRv4HM models agree across a wide range of mass ratios, this does not imply the oscillatory waveforms themselves will necessarily agree. This can be seen directly from Eq.~\eqref{eq:fluxharmonic}, which is less sensitive to small dephasing than the overlap integral commonly used to compare waveform models.

\subsection{Comparison between spin memory effects computed using different waveform models}
\label{spin_mem_comparison}

\begin{figure}[t]
	\includegraphics[scale=0.57]{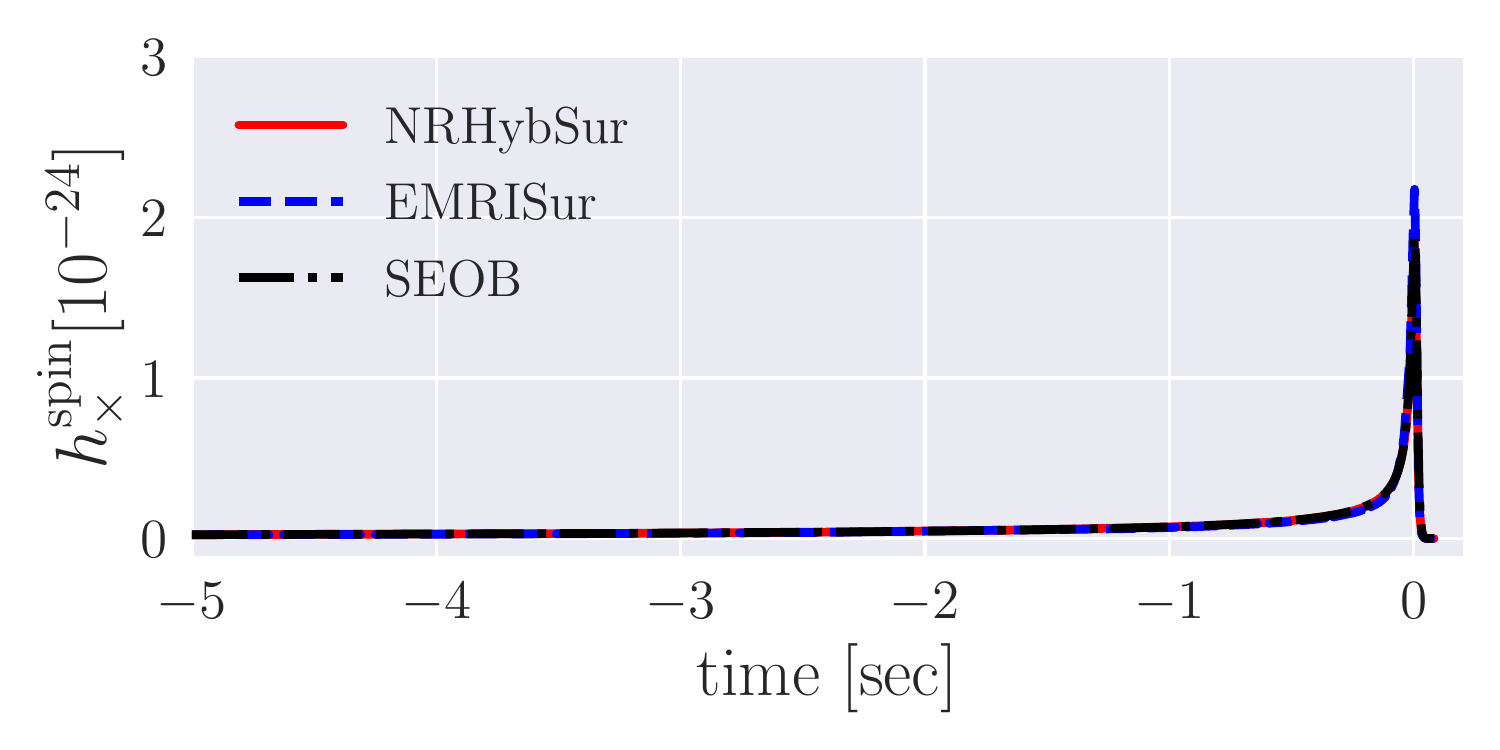}
	\caption{Spin memory computed for a non-spinning binary with mass ratio $q=10$ using three different waveform models: an aligned-spin EOB model \texttt{SEOBNRv4HM} (\textit{dashed dotted lines; labeled as} \texttt{SEOB}), an EOB-NR hybridized aligned spin surrogate waveform \texttt{NRHybSur3dq8} (\textit{solid lines; labeled as} \texttt{NRHybSur}), and a $\alpha$-calibrated ppBHPT waveform from the surrogate model \texttt{EMRISur1dq1e4}  (\textit{dashed lines; labeled as} \texttt{EMRISur}). We fix $M=200$ $M_{\odot}$, $e_{\rm ref}=0.0$ and $D=250$ Mpc.
	Relative differences between the spin memory computed using these models are always less than 15\%.
	The observed agreement between the spin memory effects computed with these three different models gives us confidence that the memory effects computed for IMRIs using either \texttt{NRHybSur3dq8}, \texttt{SEOBNRv4HM}, or the $\alpha$-calibrated \texttt{ppBHPT} waveforms accurately capture the true spin memory in this regime.
	}
	\label{Fig:spin_memory_comparison}
\end{figure}

As a final sanity check, we compute the spin memory for a binary with mass ratio $q=10$,
$M=200$ $M_{\odot}$, $e_{\rm ref}=0.0$, and $D=250$ Mpc. We restrict ourselves to a non-spinning system 
so that we can generate a high-accuracy \texttt{NRHybSur3dq8} waveform, which can be extrapolated to $q=10$
with higher accuracy provided $\chi=0$. In Fig.~\ref{Fig:spin_memory_comparison}, we show the spin memory effect computed using \texttt{SEOBNRv2}, \texttt{NRHybSur3dq8}, and the $\alpha$-calibrated ppBHPT waveform based surrogate model \texttt{EMRISur1dq1e4}. Maximum relative differences between the spin memory computed using \texttt{EMRISur1dq1e4} and \texttt{NRHybSur3dq8} (or \texttt{SEOBNRv2}) is always $\le15\%$.

\section{Phenomenology \& Detectability} 
\label{phenomenology}

In this section, we explore the memory phenomenology and detectability as the mass ratio, spin, and eccentricity is varied. 
We report SNRs computed using the design sensitivity of detectors including advanced LIGO, Cosmic Explorer (CE), and Einstein Telescope (ET). 
Assuming Gaussian detector noise, an SNR of $\approx 5$ is typically considered sufficient for detection. 
For multiple ``stacked" detections some authors have considered a total memory SNR value as low as $3$ to be sufficient for claiming hints of memory~\cite{Johnson:2018xly,Boersma:2020gxx,Lasky:2016knh}.

For most of our SNR results, we fix the intrinsic BBH parameters and luminosity distance, $D$, and report the maximum and angle-averaged SNR values. To compute the angle-averaged SNR, we simulate a total of 1125 signal realizations where we sample right accession $\alpha$ and polarization $\psi$ uniformly in $[0,2\pi]$, declination $\delta$ and inclination $\iota$ uniformly in $\cos\delta$ and $\cos\iota$ from $[-1,1]$. The maximum SNR is taken to be the largest value over the 1125 signal realizations. As the SNR is proportional to 1/D, our SNR results can easily be scaled to different luminosity distance values. Our default choice of $D=250$Mpc is motivated by the inferred distance for the event GW190814 \cite{LIGOScientific:2020zkf}, highest mass ratio event ($q\sim10$) detected by LIGO/Virgo so far.
As GW190814 is one of the closest BBH detections to date, our choice of $D=250$Mpc is represents a plausibly-optimistic default value. While we will broadly consider IMRI systems, one particular focus is on memory from hierarchical mergers involving second- or third-generation black holes. These systems present, on average, both larger masses and larger spins~\cite{gerosa2021hierarchical}. Many of our experiments consider systems with $M=200$ $M_{\odot}$, $q=10$, and large-spin systems, which is consistent with a GW190521-like remnant capturing a first-generation, stellar-mass black hole.

\begin{figure}[h]
	\includegraphics[width=\columnwidth]{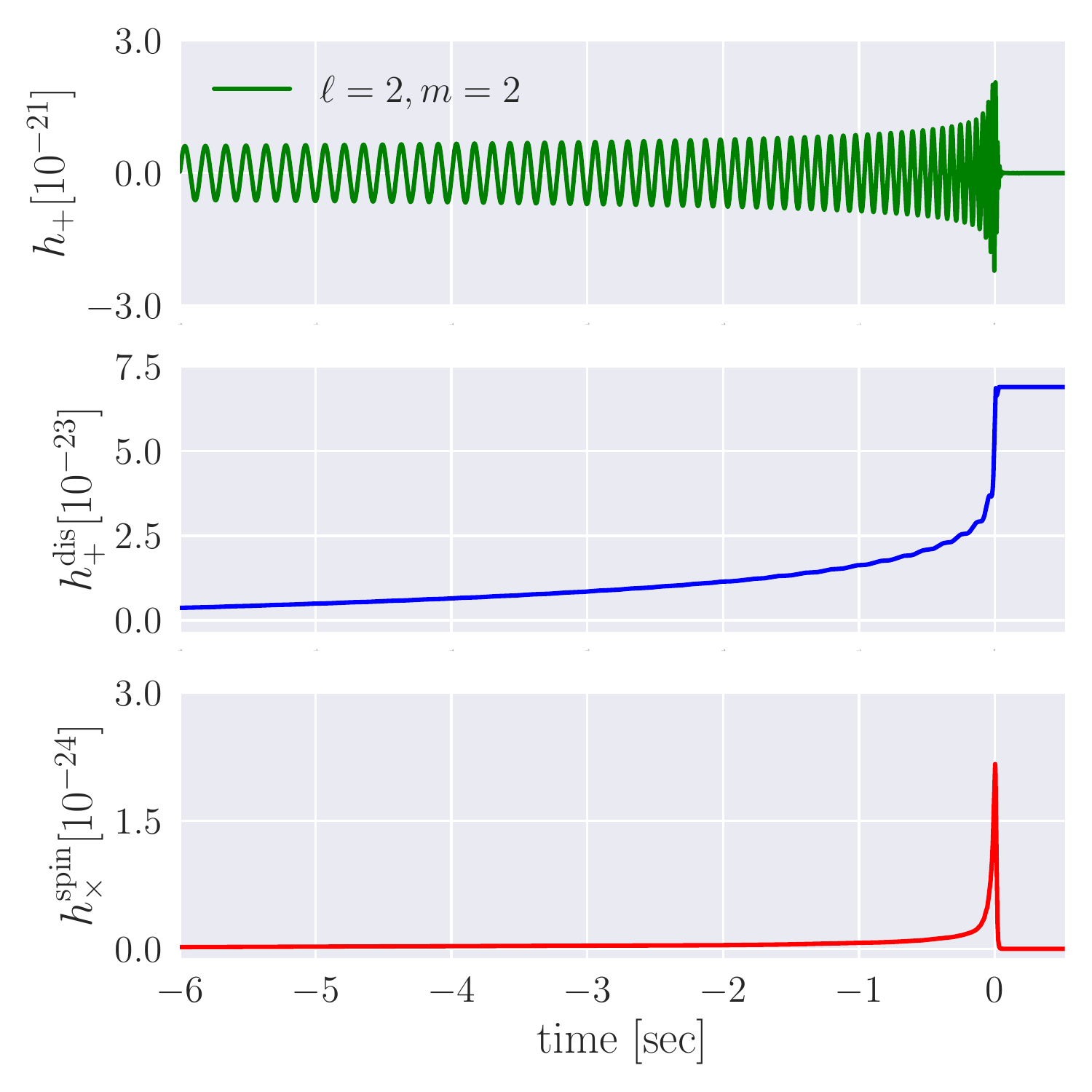}
	\caption{ \textit{Upper panel:} Plus polarization of the $h_{22}$ spherical harmonic mode for a non-spinning BBH with total mass $M=200$ $M_{\odot}$, mass ratio $q=10$ at a luminosity distance $D=250$ Mpc and at an inclination $\iota=\pi/4$.
		\textit{Middle panel:} Gravitational waveforms associated with the non-linear displacement memory contributions computed using all available modes using the higher multipole model.
		\textit{Lower panel:} Spin memory contributions computed from the dominant $\ell=2,m=2$ mode.
	}
	\label{Fig:memory}
\end{figure}

\begin{figure}[h]
	\includegraphics[width=\columnwidth]{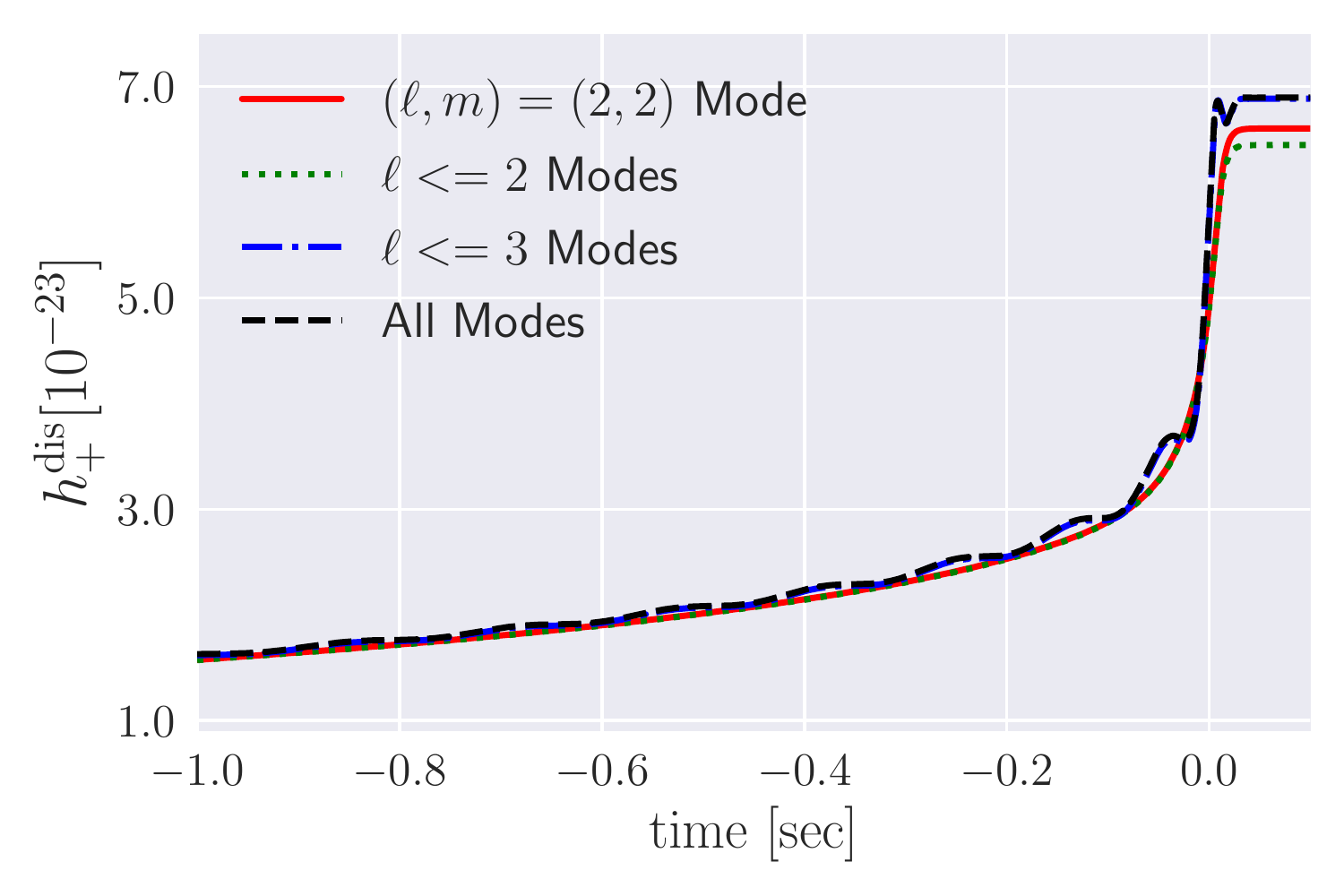}
	\caption{Gravitational waveform associated with the non-linear displacement memory contributions computed using different combinations of spherical harmonics modes. Our system is a non-spinning BBH with total mass $M=200$ $M_{\odot}$, mass ratio $q=10$ at a luminosity distance $D=250$ Mpc and at an inclination $\iota=\pi/4$. We zoom in the late inspiral, merger and ringdown part of the waveform. 
	}
	\label{Fig:memory_mode}
\end{figure}

\subsection{Structure of the memory signal}
\begin{figure}[t]
	\includegraphics[width=\columnwidth]{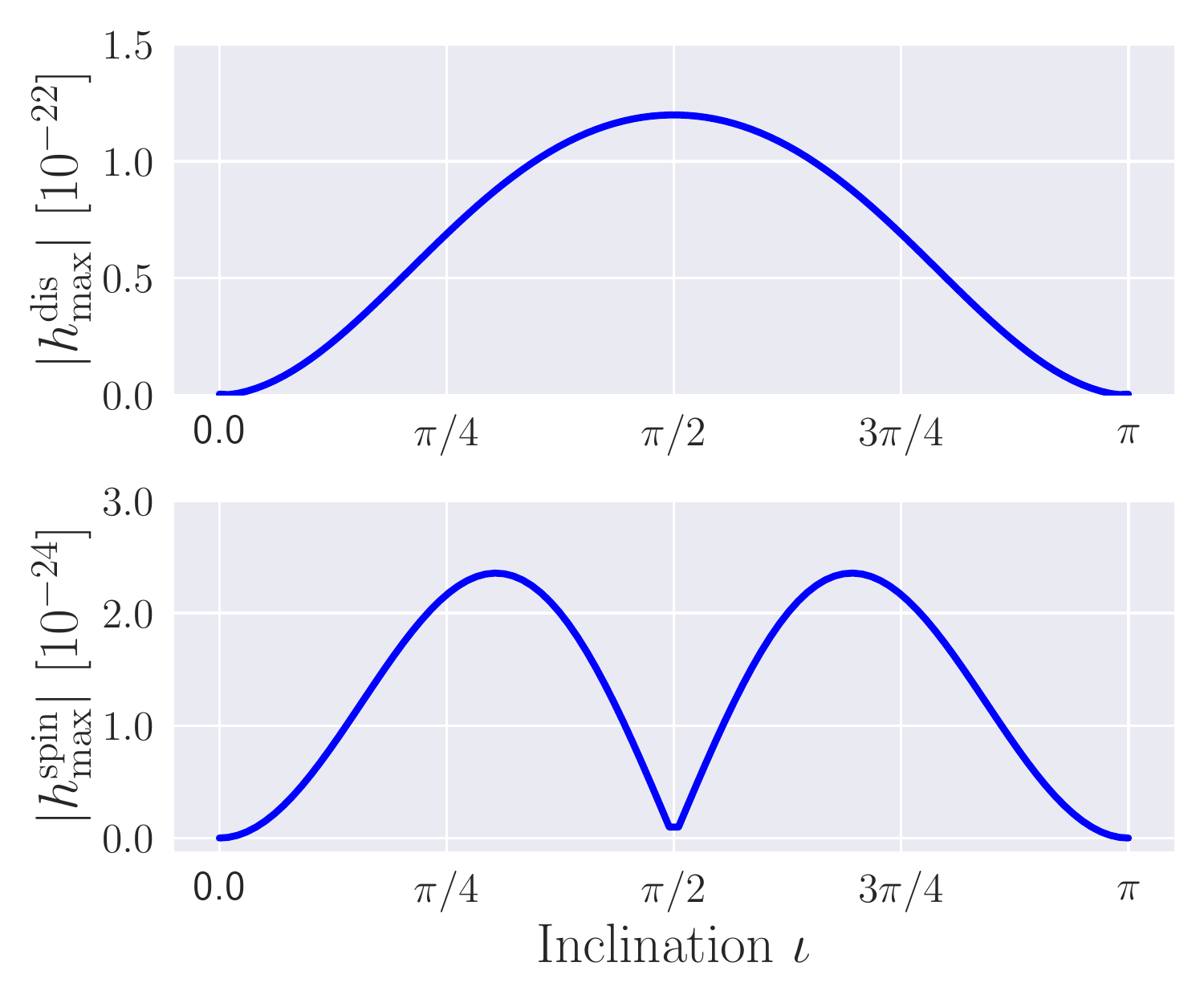}
	\caption{Maximum of the total displacement memory and spin memory as a function of the inclination angle $\iota$. Our system is a non-spinning BBH with total mass $M=200$ $M_{\odot}$, mass ratio $q=10$ at a luminosity distance $D=250$ Mpc.
		\label{Fig:memory_iota}
	}
\end{figure}

To understand the structure of memory waveforms, we pick a non-spinning GW signal with mass ratio $q=10$, total mass $M=200$ $M_{\odot}$ and luminosity distance $D=250$ Mpc. 
In Fig.~\ref{Fig:memory}, we plot both the displacement memory (\textit{middle panel}) and spin memory (\textit{lower panel}) contributions from the dominant $\ell=2,m=2$ mode. 
For comparison, we also show $\ell=2,m=2$ mode waveform (\textit{upper panel}). 
We fix the inclination to be $\iota=\pi/4$ such that the memory effect fits in between the maximum and conservative cases (discussed more in Fig.~\ref{Fig:memory_iota}).
Displacement memory effects are found to be two orders of magnitude smaller than the oscillatory waveform,
and the spin memory contributions are another $\sim$ two orders of magnitude smaller compared to its displacement memory counterpart. We note that the displacement memory increases gradually during the inspiral and reaches a flat maximum following the merger. Spin memory, on the other hand, drops sharply after the merger. 

In Table~\ref{tab:1} we report the maximum and angle-averaged SNR for a BBH with $q=10$, $M=200$ $M_{\odot}$ and $D=250$ Mpc in four different detectors. We find that while displacement memory modes will have significant SNRs in future detectors and could possibly result in confident detection, spin memory modes would still have very low SNRs. 
These SNR values are consistent with earlier studies done in the context of comparable mass binaries \cite{Johnson:2018xly,Nichols:2017rqr,McNeill:2017uvq}. 

To probe the effects of higher modes, in Fig.~\ref{Fig:memory_mode} we show the total memory computed using different mode content for a binary with $q=10$ and $M=200M_{\odot}$. 
To compute the displacement memory we use 
(i) only the dominant $(2,\pm2)$ modes (solid red line), 
(ii) all modes with $\ell\le2$ (dashed green line), 
(iii) all modes with $\ell\le3$ (dash-dot blue line), 
(iv) and all modes in our $\alpha$-calibrated ppBHPT waveforms (dashed black line).
We notice that the memory contribution from the quadrupolar mode
already accounts for most of the signal content.
To quantify the importance of the higher modes in the memory computation we compute SNRs of displacement memory signals obtained using only the quadrupolar mode (Table \ref{tab:1}; in parenthesis). We find that SNRs increase by about $\sim 7\%$ across detectors when higher order modes are included in memory computation.  

We now probe the dependence of both the displacement and spin memory on the inclination angle $\iota$. 
In Fig.~\ref{Fig:memory_iota}, we plot the maximum of the displacement and spin memory computed using all available modes as a function of the $\iota$. 
We fix $q=10$, $M=200$ $M_{\odot}$, and $D=250$ Mpc. 
We find that the maximum effects for the displacement memory is obtained for $\iota=\pi/2$ whereas spin memory modes are loudest for $\iota \sim \pi/4 - \pi/3$. 
This is due to the fact that for the displacement memory the $(2,0)$ mode is dominant over other modes, and the angular
dependency of a waveform consisting of the $(2,0)$ mode only is proportional to $\sim\sin^2\iota$ (as done in quadrupole approximation).

\begin{table}
	\caption{Angle-averaged and maximum SNRs~\footnote{Symbols: $\rho_{\rm avg}$: average SNR; $\rho_{\rm max}$: maximum SNR.} of the displacement memory mode and spin memory mode in different detectors (aLIGO, KAGRA \cite{Somiya:2011np}, ET and Cosmic Explorer (CE) \cite{Evans:2016mbw}). For all detectors, we use their design sensitivity. The BBH source parameters are: $q=10$, $M=200$ $M_{\odot}$, $D=250$ Mpc.
	For a comparison we also show the SNRs (in parenthesis) for memory signals computed using only the dominant $l=2,m=\pm2$ mode.
	}
	\begin{ruledtabular}
		\begin{tabular}{l | c c | c c}			
			&Displacement Memory & & Spin Memory\\ 
			\hline
			&$\rho_{\rm avg}$ &$\rho_{\rm max}$ &$\rho_{\rm avg}$ &$\rho_{\rm max}$\\
			\hline
			aLIGO &0.39 &1.69 &0.001 &0.009	\\
			&(0.38) &(1.63) & &	\\
			\hline
			KAGRA 	&0.29 &1.24 &0.001 &0.004 \\
			&(0.28) &(1.19) & &	\\
			\hline
			Virgo &0.316 &1.367 &0.001 &0.005 \\
			&(0.308) &(1.194) & & \\
			\hline 
			ET	&4.46 &19.29 &0.03 &0.19\\
			&(4.30) &(18.42) & &	\\
			\hline
			CE  &14.91 &64.85 &0.12 &0.81\\
			&(14.18) &(60.748) & &	\\
		\end{tabular}
	\end{ruledtabular}
	\label{tab:1}
\end{table}

\begin{figure*}[t]
	\includegraphics[scale=0.5]{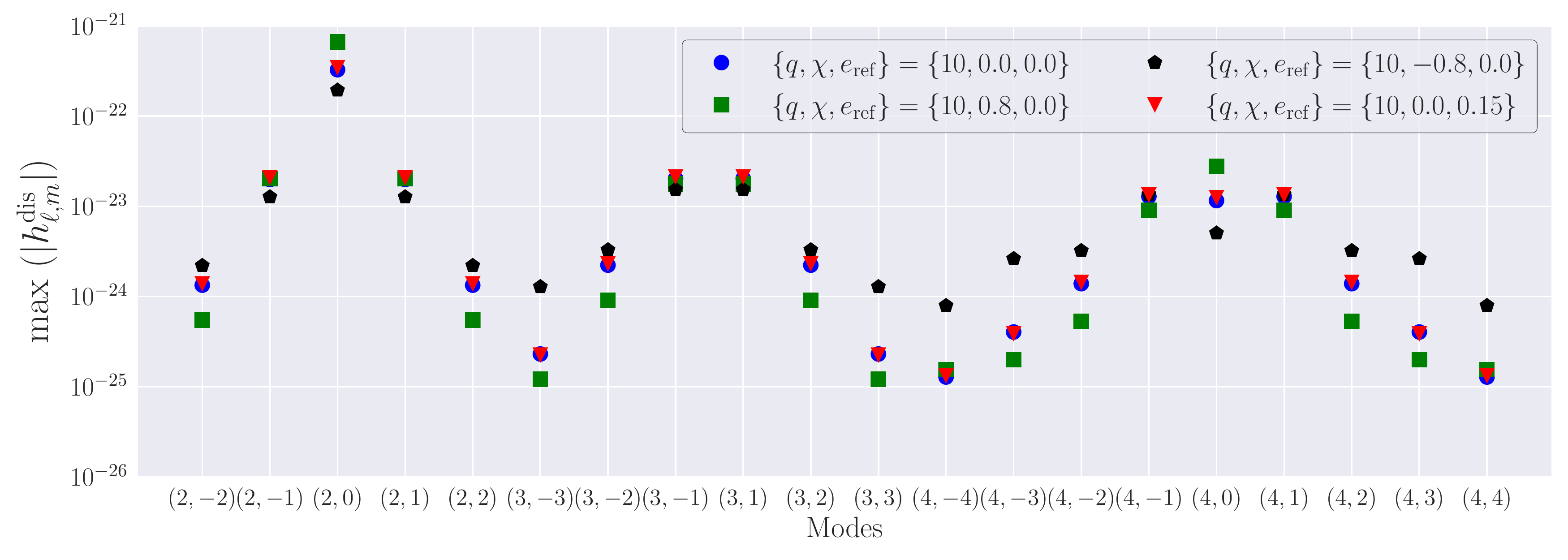}
	\caption{The spherical harmonic decomposition of the displacement memory waveform for different spin and eccentric configurations. We fix mass ratio $q=10$, $M=200$ $M_{\odot}$ and $D=250$ Mpc. The absolute value of the late-time  memory  is  shown  as  a  function  of  the  $(\ell,m)$  spherical harmonic decomposition of the memory. This figure extends Fig 3 of Talbot et al.~\cite{Talbot:2018sgr} in the high mass ratio regime - focusing on mass ratio $q=10$ and for different configurations of spins and eccentricities. The (3,0) mode's amplitude is extremely small and omitted from this figure. \label{Fig:memory_modes_q10}
	}
\end{figure*}

\begin{figure*}[t]
	\includegraphics[scale=0.5]{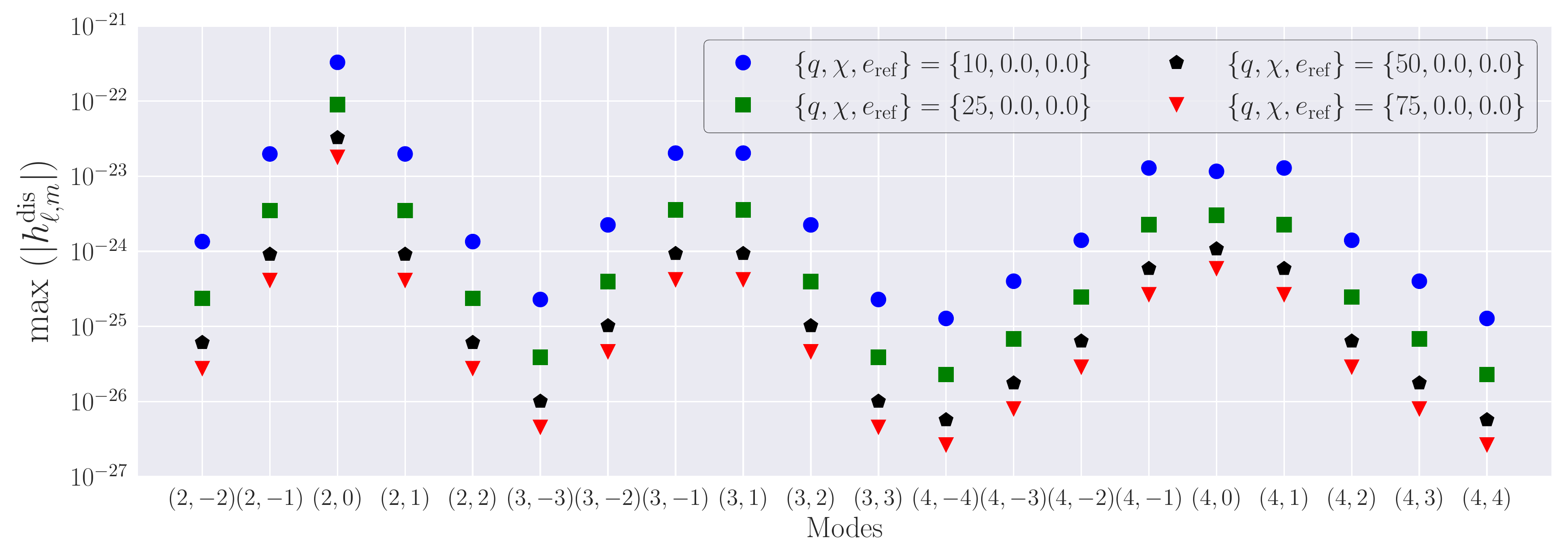}
	\caption{The spherical harmonic decomposition of the displacement memory waveform for different mass ratios. We fix spin $\chi=0.0$, $e_{\rm ref}=0.0$, $M=200$ $M_{\odot}$ and $D=250$ Mpc. The absolute value of the late-time  memory  is  shown  as  a  function  of  the  $(\ell,m)$  spherical harmonic decomposition of the memory. 
	This figure extends Fig 3 of Talbot et al.~\cite{Talbot:2018sgr} in the high mass ratio regime - covering
	mass ratios $10\le q \le 75$.  The (3,0) mode's amplitude is extremely small and omitted from this figure.
		\label{Fig:memory_modes_q1075}
	}
\end{figure*}

\subsection{Mode Decomposition of the Memory Waveform}
Following the prescription provided in Ref.~\cite{Talbot:2018sgr}, we decompose the memory waveform into spin-weighted harmonics modes to explore their dependence on spin and eccentricity (Fig.~\ref{Fig:memory_modes_q1075}), and mass ratio (Fig.~\ref{Fig:memory_modes_q10}). In all cases, as is well known, we find that the $(2,0)$ mode is dominant. Many of the subdominant modes (except the $(2,1)$ and $(3,1)$) are negligible compared to the $(2,0)$ mode and could safely be ignored for SNR computations, although we will continue to include them. 
We further find that $(\ell,m)$ and $(\ell,-m)$ memory modes have exactly same maximum amplitudes, as we would expect for systems that obey orbital-plane symmetry.

Fig.~\ref{Fig:memory_modes_q10} shows the mode decomposition of $q=10$ binaries while varying spin and eccentricity configurations. By comparing the quasi-circular (blue circle) and eccentric (red triangle) cases, we see that eccentricity brings almost no change in the maximum value of the memory modes. This has been observed for the dominant $(2,0)$ memory mode in the context of comparable mass ratio binaries \cite{Favata:2011qi,Liu:2021zys}, and our result extends this finding in the intermediate mass ratio regime and for the subdominant memory modes. Spinning systems, however, have noticeably different mode content as compared to their non-spinning counterparts; Sec.~\ref{sec:snr_spin} considers the impact of spin on the memory's SNR.

Fig.~\ref{Fig:memory_modes_q1075} shows the different memory modes for non-spinning, non-eccentric binaries as the mass ratio is increased. The effect of mass ratio is clearly observed in Fig.~\ref{Fig:memory_modes_q1075} where the maximum value in all of the memory modes decreases with mass ratio. Ref.~\cite{Talbot:2018sgr}, however, observed an increased contribution to higher order modes from the $q \leq 2$ asymmetric mass binaries they considered, which is also apparent in Fig.~\ref{Fig:memory_modes_NRHybSur3dq8}.


\begin{figure}[t]
	\includegraphics[width=\columnwidth]{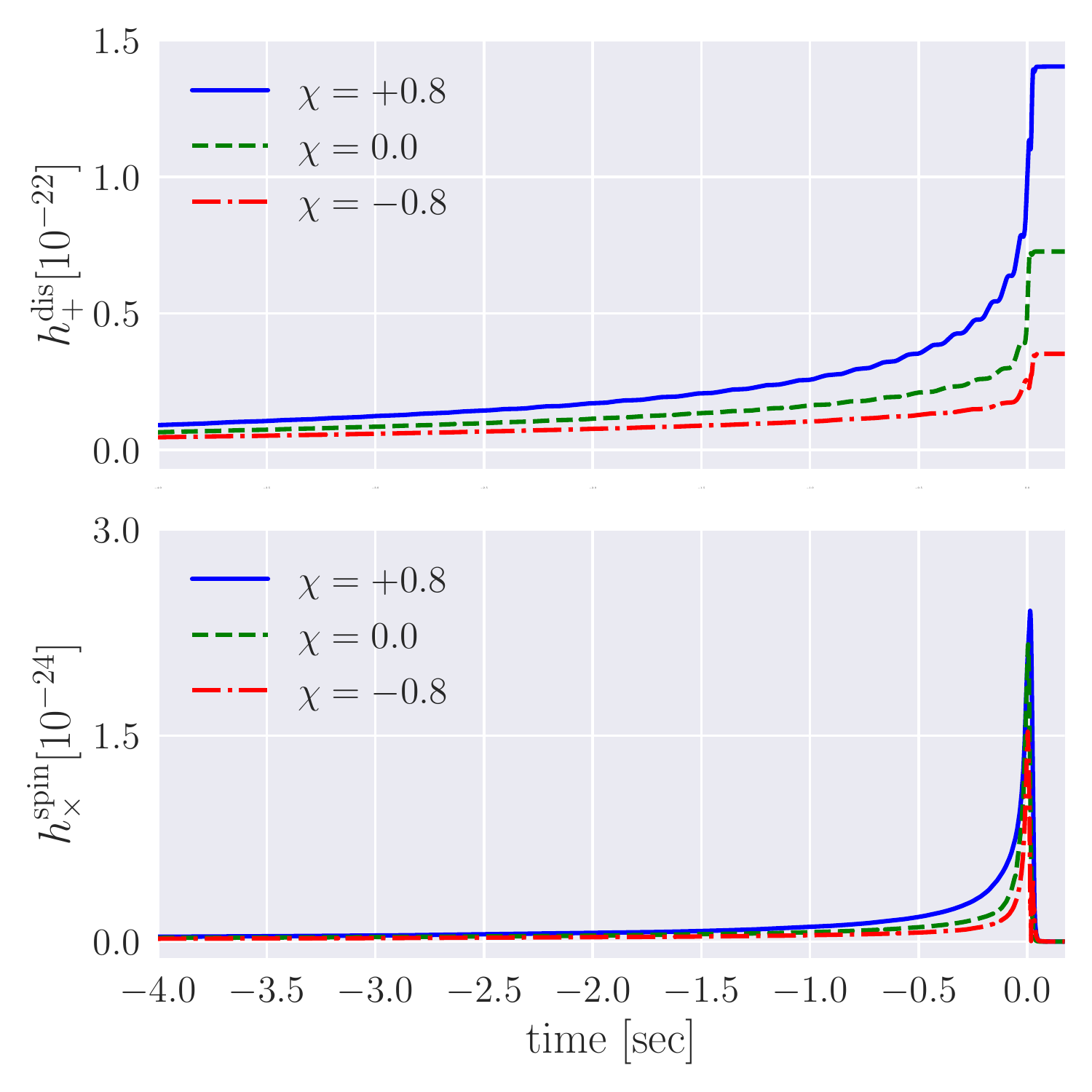}
	\caption{Time evolution of the displacement memory $h^{\rm dis}(t)$ \textit{(upper panel)} and spin memory $h^{\rm spin}(t)$\textit{(lower panel)} for three different spin values, $\chi$, of the primary black hole. Our system is a BBH with total mass $M=200$ $M_{\odot}$, mass ratio $q=10$ at a luminosity distance $D=250$Mpc and inclination $\iota=\pi/4$. 
		\label{Fig:memory_spin}
	}
\end{figure}

\begin{figure}[t]
	\includegraphics[width=\columnwidth]{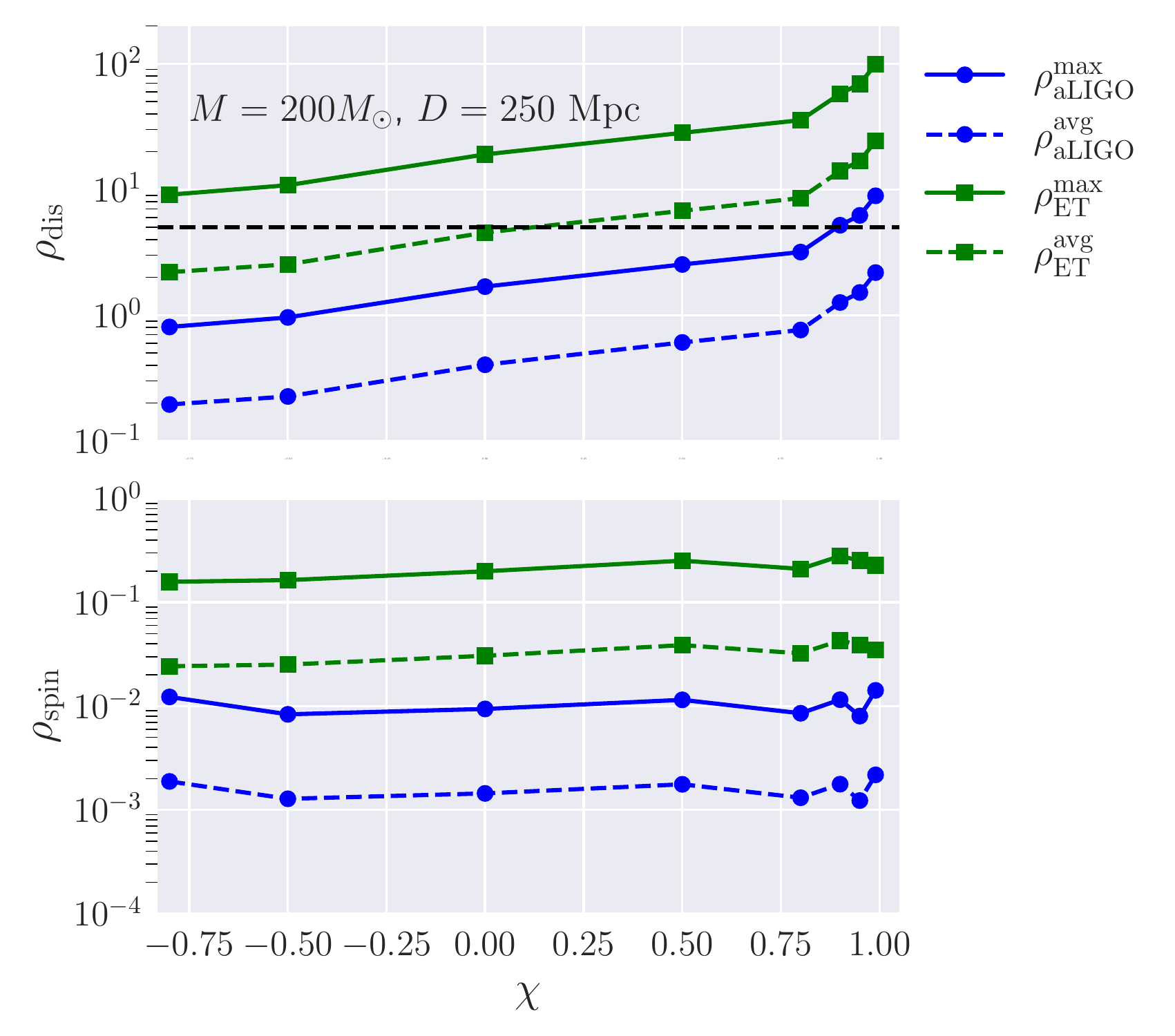}
	\caption{Maximum SNR $\rho_{\rm max}$ (solid line) and angle-average SNR $\rho_{\rm avg}$ (dashed line) for the memory modes computed using the design sensitivity of the advanced LIGO (blue) and ET (green) as a function of the spin on the primary black-hole. We use $q=10$, $M=200$ $M_{\odot}$ and $D=250$ Mpc. Upper panel (lower panel) shows the SNR for the displacement memory (spin memory). Black dashed line denotes an SNR of 5, typical threshold for detection.
		\label{Fig:memory_spin_snr}
	}
\end{figure}
\subsection{Effect of spin} \label{sec:snr_spin}

Next, we provide a systematic study of both displacement and spin memory's spin dependence. We compute the memory effects for a set of binaries with mass ratio $q=10$ but for different values of the primary black-hole's dimensionless spin $\chi$. We fix $q=10$, $M=200$ $M_{\odot}$, $D=250$ Mpc and $\iota=\pi/4$. 
In Fig.~\ref{Fig:memory_spin}, we show the displacement and spin memory effects as a function of time for three different BBH with spins $\chi=[-0.8,0.0,0.8]$. We observe that the memory effect increases as the spin $\chi$ increases. This is due to the fact that prograde inspiral spends a longer in the strong field as the last stable orbit's radius shrinks. As a consequence, the SNR of emitted GWs is also expected to be larger than the corresponding non-spinning binary system. Our findings are consistent with results obtained in Ref.~\cite{Burko:2020gse} that observes an increased memory amplitude for lager values for spins.

In Fig.~\ref{Fig:memory_spin_snr}, we report the maximum and angle-averaged SNR for the memory modes in a BBH with $q=10$, $M=200$ $M_{\odot}$ and $D=250$ Mpc. The maximum SNRs for the displacement memory in ET is sufficient for confident detection across the entire range of spins considered here. The angle-averaged SNR for the displacement memory is between $\sim 1$ (for $\chi=-.8$) and $\sim 20$ (for $\chi=.99$). For hierarchical mergers with second-generation (or higher) component black holes, spins of up to $\chi\approx.9$ are expected~\cite{doctor2020black,gerosa2021hierarchical}, suggesting that memory from some of these system may be directly observed with ET even in the typical case. Indeed, a handful of high-spin black holes have been identified \cite{Zackay:2019tzo,Qin:2018sxk} and may therefore be the most promising candidates for memory detections. For aLIGO, angle-averaged SNRs are always below the value 5 at our fiducial distance $D=250$ Mpc, although optimally oriented binaries cross the detection threshold for $\chi>=0.9$. Increased SNR at larger spin values would therefore favorably contribute to forecasts for memory detection. The spin memory, however, is unlikely to be detected in aLIGO or ET.

\begin{figure}[t]
	\includegraphics[width=\columnwidth]{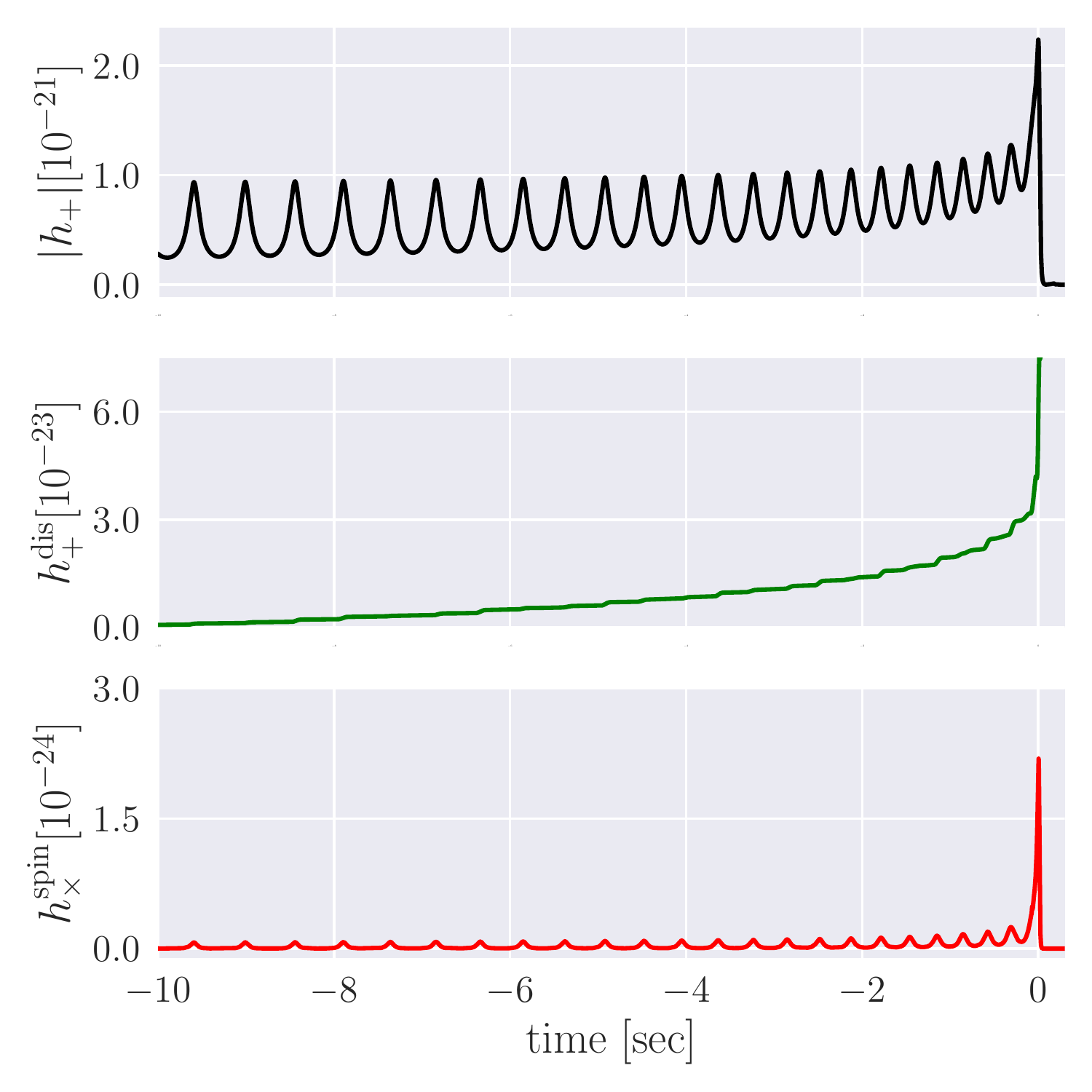}
	\caption{Time evolution of the displacement memory $h^{\rm dis}(t)$ \textit{(middle panel)} and spin memory $h^{\rm spin}(t)$ \textit{(lower panel)} for the BBH with eccentricity $e_{\rm ref}=0.17$ measured three cycles before the merger. All other details are same as in Fig.~\ref{Fig:memory}. For a comparison, we show the time evolution of the $\ell=2,m=2$ mode amplitude in the upper panel. Eccentricity introduces additional modulation in both the amplitude of $\ell=2,m=2$ oscillatory mode and both flavors of memory.
		\label{Fig:memory_ecc}
	}
\end{figure}

\begin{figure}[t]
	\includegraphics[width=\columnwidth]{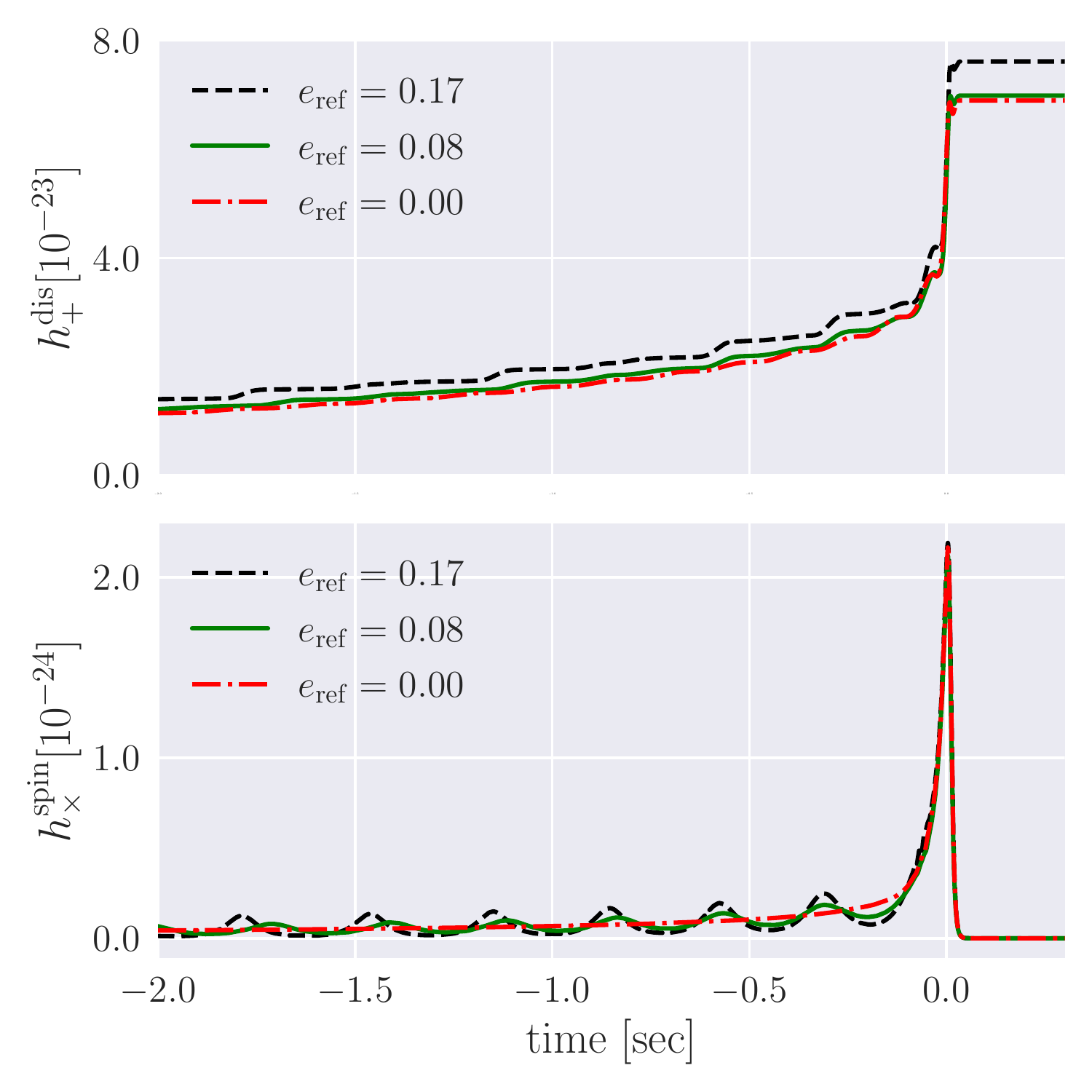}
	\caption{Time evolution of the displacement memory $h^{\rm dis}(t)$ \textit{(upper panel)} and spin memory $h^{\rm spin}(t)$ \textit{(lower panel)} for three different eccentricity values $e_{\rm ref}$. Our system is a BBH with total mass $M=200$ $M_{\odot}$, mass ratio $q=10$ at a luminosity distance $D=250$ Mpc and inclination $\iota=\pi/4$.
	\label{Fig:memory_diffecc}
	}
\end{figure}

\subsection{Effect of eccentricity} \label{sec:eccentricity}

Depending on the formation channel, some IMRIs are expected to retain significant eccentricity even at the final stage of the inspiral~\cite{Rodriguez:2017pec,Samsing:2017xmd,AmaroSeoane:2007aw}.
Memory from eccentric systems has typically been studied using PN approximations~\cite{Favata:2011qi, Ebersold:2019kdc} or using a kludge model~\cite{Burko:2020gse}.
In this paper we use the higher multipole model and focus on small to moderate eccentricities with $e\lesssim 0.2$. To estimate the eccentricity at a given time, we use~\cite{Mora:2002gf}:
\begin{equation}
	\label{Eq:ecc_estimator}
	e(t)=\frac{\sqrt{\omega_p(t)} - \sqrt{\omega_a(t)}}{\sqrt{\omega_p(t)} + \sqrt{\omega_a(t)}},
\end{equation}
where $\omega_a$ and $\omega_p$ are the orbital frequencies at apocenter and pericenter, respectively. 
We let $e_{\rm ref}$ be the value of $e(t)$ measured three cycles before the merger.
A detailed description of the method is given in Refs.~\cite{Mora:2002gf,Islam:2021mha}. 

We explore how the memory effect changes as the binary becomes increasingly more eccentric. 
To do this, we simulate gravitational waveforms for $q=10$, $M=200$ $M_{\odot}$, $D=250$ Mpc, and 
$\iota=\pi/4$ with different values of eccentricity. 
In Fig.~\ref{Fig:memory_ecc}, we show the memory contributions as a function of time for one
particular value of eccentricity $e_{\rm ref}=0.17$.
Eccentricity introduces additional modulation in both displacement and spin memory components. 
Such modulations are small in the displacement memory but prominent for the spin memory. 
The modulations are strongly correlated to the modulations in the oscillatory gravitational waveform (upper panel; Fig.~\ref{Fig:memory_ecc}) and these features become more evident as the eccentricity increases (Fig.~\ref{Fig:memory_diffecc}). 
For the displacement memory, these modulations roughly resembles the staircase structure found in the zoom-whirl orbits \cite{Burko:2020gse}. These modulations are consistent with results obtained in Refs. \cite{Favata:2011qi,Liu:2021zys} for comparable mass ratio binaries.

Next, we compute the SNRs for memory signals from eccentric binaries. 
In Table \ref{tab:2} we report SNRs for the binary with highest eccentricity ($e_{\rm ref}=0.17$) considered in our study. 
For comparison, we also show the SNR values computed for the non-eccentric binary.
The SNR values change by at most 4 percent.
We find that for a given mass ratio and detector sensitivity, the computed SNRs are roughly constant for $e_{\rm ref}\leq0.17$ despite the rich phenomenology that larger eccentricity offers.
This is perhaps not surprising in light of the fact that most of the SNR is accumulated around the merger, and maximum value for the memory signal changes modestly with eccentricity for the values of $e_{\rm ref}$ considered here.
We further confirm that the differences in SNR computed using ppBHPT waveforms with and without $\alpha$ scaling [defined in Eq.~(\ref{alpha})] are small, suggesting that the rescaling obtained from the non-eccentric binaries can reasonably be used for eccentric binaries - at least for the purpose of this study.

\begin{figure}[t]
	\includegraphics[width=\columnwidth]{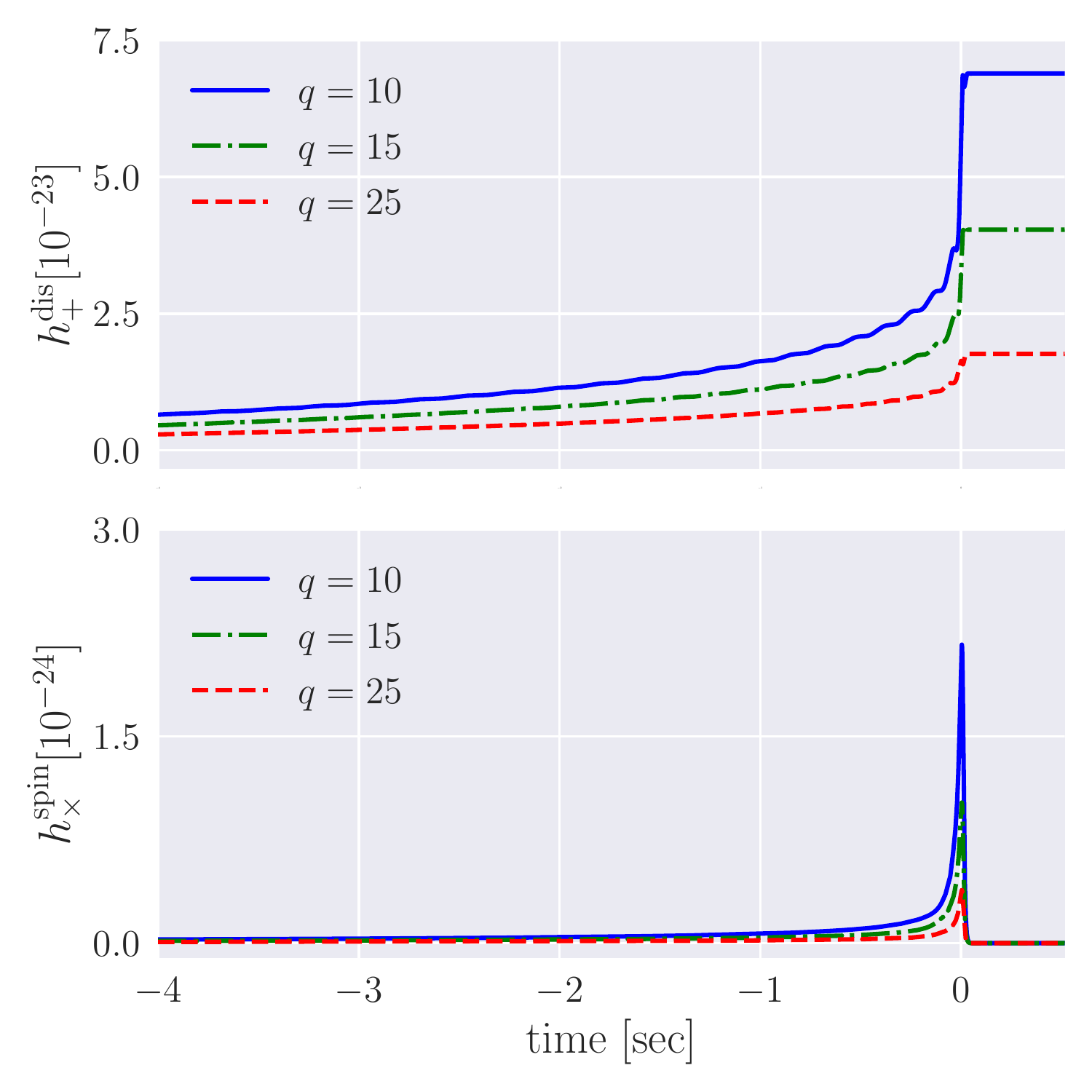}
	\caption{Time evolution of the displacement memory $h^{\rm dis}(t)$ \textit{(upper panel)} and spin memory $h^{\rm spin}(t)$ \textit{(lower panel)} for BBHs with three different mass ratio $q$. Our system is a non-spinning BBH with total mass $M=200$ $M_{\odot}$, at a luminosity distance $D=250$ Mpc and at an inclination $\iota=\pi/4$.
		\label{Fig:memory_diffq}
	}
\end{figure}

\begin{figure}[t]
	\includegraphics[width=\columnwidth]{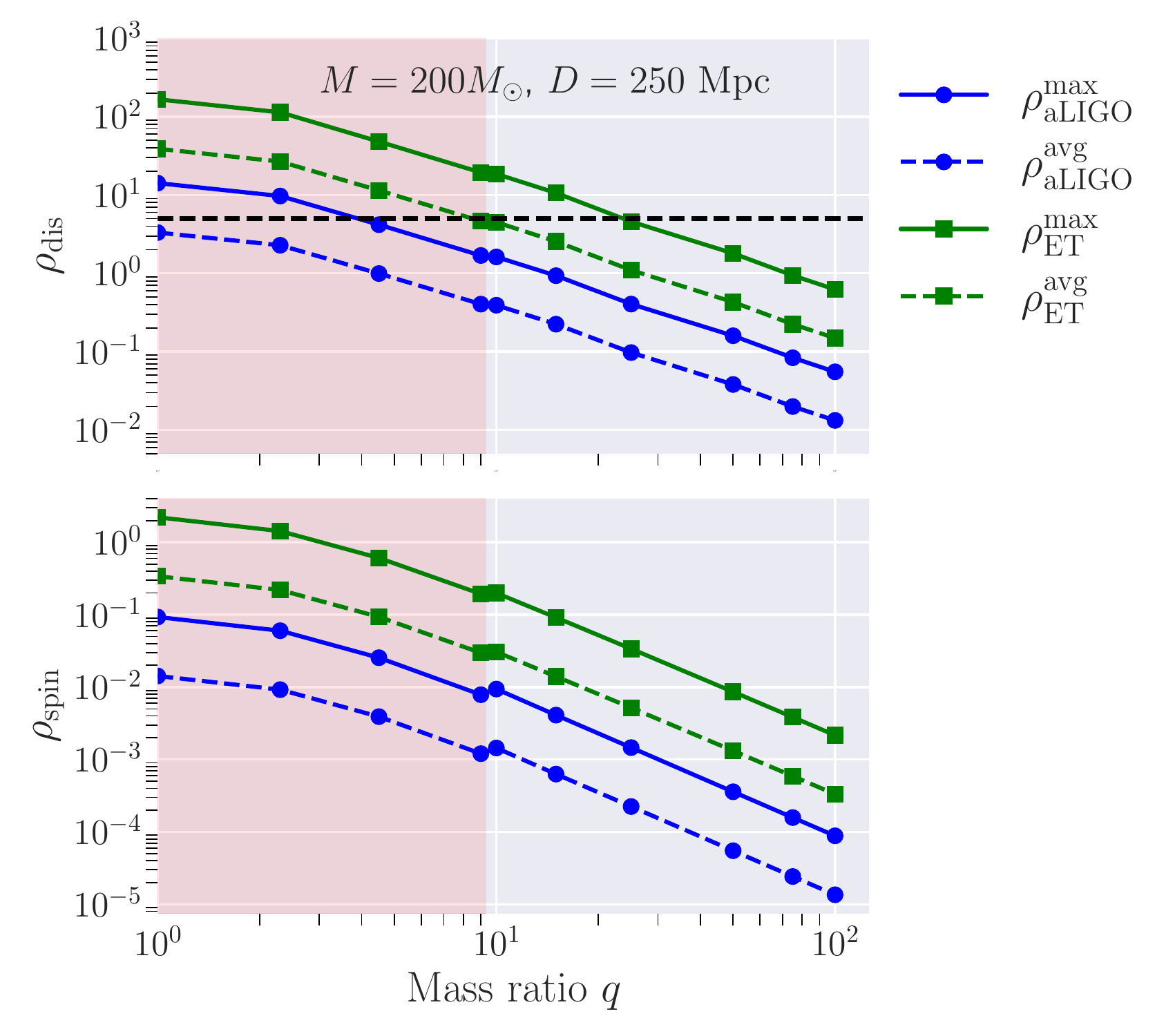}\label{Fig:memory_diffq_snr}
	\caption{Maximum SNR $\rho_{\rm max}$ (solid line) and angle-average SNR $\rho_{\rm avg}$ (dashed line) for the memory modes computed using the design sensitivity of the advanced LIGO (blue) and ET (green) as a function of the mass ratio $q$. We use $M=200$ $M_{\odot}$, $\chi=0.0$, $e_{\rm ref}=0.0$ and $D=250$ Mpc. Both displacement and spin memory decreases as $\sim 1/q$.
	The shaded red region shows SNRs computed using the \texttt{NRHybSur3dq8} model over $1\le q \le 9$ (details are in text).
	}
\end{figure}

\begin{table}
	\caption{Angle-averaged and maximum SNRs of the displacement memory mode and spin memory mode in different detectors for 
	a $M=200$ $M_{\odot}$, $\chi=0.0$, $D=250$ Mpc binary with highest eccentricity ($e_{\rm ref}=0.17$) considered in our study.  For all detectors, we use their design sensitivity.
	For a comparison, we also show the SNRs of the corresponding non-eccentric binary in parenthesis.
	}
	\footnote{Symbols: $\rho_{\rm avg}$: average SNR;  
		$\rho_{\rm max}$: maximum SNR.
	}
	\begin{ruledtabular}
		\begin{tabular}{l | r l | r l}			
			&Displacement &Memory &Spin &Memory \\ 
			\hline
			&$\rho_{\rm avg}$ &$\rho_{\rm max}$ &$\rho_{\rm avg}$ &$\rho_{\rm max}$\\
			\hline
			aLIGO &0.409 &1.772&0.001 &0.010	\\
			      &(0.39) &(1.69) &(0.001) &(0.009)	\\
			\hline
			KAGRA &0.298 &1.290 &0.001 &0.004 \\
			      &(0.29) &1.235 &0.001 &0.004 \\
			\hline
			Virgo &0.330 &1.427 &0.001 &0.005 \\
			      &(0.316) &(1.367) &(0.001) &(0.005) \\
			\hline
			ET	&4.64 &20.140 &0.031 &0.201 \\
			    &(4.46) &(19.295) &(0.031) &(0.199) \\
			\hline
			CE  &15.516 &67.632 &0.125 &0.817\\
			    &(14.915) &(64.848) &(0.124) &(0.810)\\
		\end{tabular}
	\end{ruledtabular}
	\label{tab:2}
\end{table}

\subsection{Effect of mass ratio} 
We explore how the memory effect changes as the binary becomes increasingly asymmetric. To do this, we simulate 
gravitational waveforms with $M=200$ $M_{\odot}$, $\chi=0.0$, $e_{\rm ref}=0.0$, 
$D=250$ Mpc and $\iota=\pi/4$ while varying the mass ratio.
In Fig.~\ref{Fig:memory_diffq}, we plot both the displacement memory and spin memory for three different mass ratios $q=\{10,15,25\}$. 
The memory signals become weaker as the mass ratio increases, which follows 
from the fact that the oscillatory mode's amplitude decreases as $1/q$ in the large-mass-ratio limit. 
In Fig.~\ref{Fig:memory_diffq_snr}, we show the maximum and angle-averaged SNR for memory signals with different mass ratios. We find that the SNRs for the memory modes decrease as mass ratio increases for $q \geq 10$.
We complement this analysis by computing SNRs for the memory signals in the comparable mass ratio regime ($1\le q \le9$) using the \texttt{NRHybSur3dq8} model (red shaded region). We confirm that the overall trend of SNR scaling ($\sim1/q$) continues almost up to $q=1$. Furthermore, we observe that
the SNR values have a mostly smooth transition from the comparable mass regime (obtained using \texttt{NRHybSur3dq8} model) to intermediate mass ratio regime (obtained using $\alpha$-scaled ppBHPT waveforms) providing further evidence on the robustness of the SNR computation in the $q \geq 10$ regime where waveform modeling is more challenging.

\begin{figure}[t]
	\includegraphics[scale=0.62]{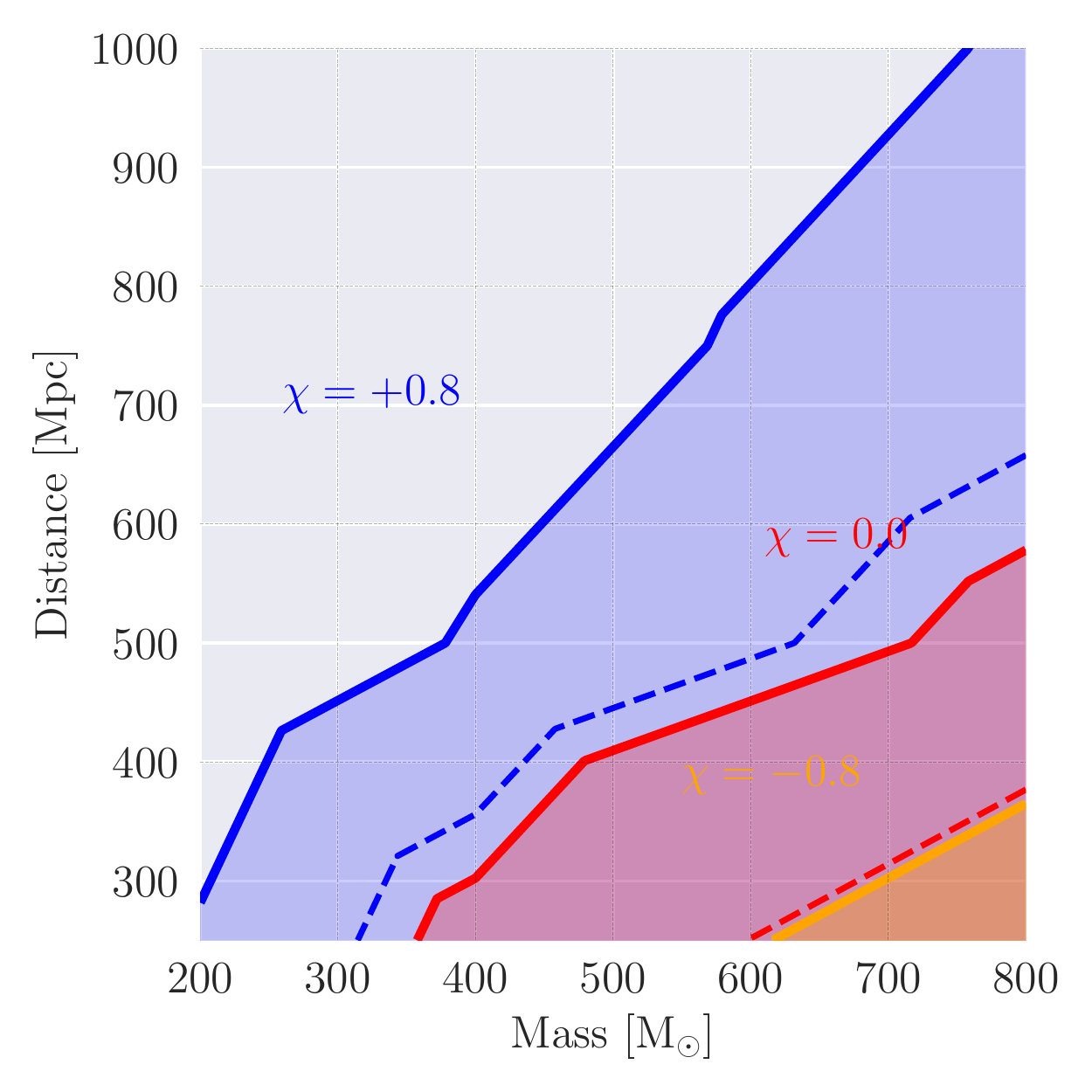}
	\caption{We show the region (shaded in blue/red/orange) in parameter space (for a $q=10$ binary) important for the detection of displacement memory with a single advanced LIGO detector operating at design sensitivity. The solid red line denotes the total mass and distances for which an optimally-oriented nonspinning ($\chi=0.0$) binary has an SNR=3 (``hints of memory"). The red shaded area below this line indicates the region where the memory modes have SNR$\geq 3$. We also show similar boundaries for $\chi=0.8$ (solid blue line) and $\chi=-0.8$ (solid orange line). Dashed lines are used to mark the location of SNR$=5$.
	}
	\label{Fig:q10_snrsummary}
\end{figure}

\subsection{Detectability in aLIGO}
In the previous subsections, we have focused on the dependence of the memory's SNR (with and without subdominant modes) as the total mass, spin, mass ratio, and eccentricity are varied. We now consider the regions of the mass-distance parameter space that are most promising for the direct detection of displacement memory. In light of Fig.~\ref{Fig:memory_diffq_snr}, we again focus on $q=10$ systems as they offer the best chance for direct detection for the mass ratios primarily considered in this paper.

In Fig.~\ref{Fig:q10_snrsummary}, we map out regions of plausible detectability for binaries with mass ratio $q=10$.
The shaded areas below each solid line indicate the region where displacement memory modes have a maximum SNR (optimally oriented with respect to the detector) of more than 3 (solid line; for ``hints" of memory) and 5 (dashed line).
We show boundaries for binaries with $\chi=0.8$ (blue), $\chi=0.0$ (red), and $\chi=-0.8$ (orange). With larger positive spins, displacement memory becomes increasingly easier to detect. These are particularly promising as the primary black hole of an IMRI system is generally expected to have large positive spins ($\sim 0.7$) when formed through hierarchical mergers \cite{Gerosa:2017kvu,Fishbach:2017dwv,Berti:2008af}. 
	
\section{Discussion \& Conclusion} 
\label{Sec:Conclusion}

In this work, using a recently-developed spin~\cite{Nichols:2017rqr} and higher multipole displacement memory model~ \cite{Talbot:2018sgr}, we systematically investigate the total memory effects for intermediate mass ratio inspirals (IMRIs) while primarily focusing on the potential detectability of these signals. Our work is motivated by binary systems formed through hierarchical mergers~\cite{doctor2020black,gerosa2021hierarchical}, for example, when a GW190521-like remnant captures a stellar-mass black hole. Such systems typically have a large total mass, large spin on the primary, and possibly residual eccentricity; features that potentially raise the prospect for memory detection especially when subdominant modes are included into the analysis.

To generate the oscillatory part of the IMRI waveform (which is used in the computation of memory), we use point particle black hole perturbation theory (ppBHPT) waveforms computed by solving the Teukolsky equation. The ppBHPT waveforms are then calibrated to NR simulations for $q\le 10$ using a rescaling discussed in Sec.~\ref{imridata}. 
As IMRI waveform models are still under active development, we have furnished extensive comparisons (Figures \ref{Fig:memory_modes_NRHybSur3dq8}, \ref{Fig:memory_modes_SEOB}, \ref{Fig:q10_SEOB}, and \ref{Fig:spin_memory_comparison}) between a hybrid EOB-NR surrogate \texttt{NRHybSur3dq8}, an aligned-spin effective one body model \texttt{SEOBNRv4HM}, and our calibrated ppBHPT waveforms. We find these models agree surprisingly well for the dominant contributions to the GW memory for mass ratios $q \leq 100$ despite being calibrated to NR simulations mostly in the comparable mass ratio regime. Sec.~\ref{robustness_study} provides evidence that the results presented in our paper serve as a useful and reliable ballpark estimate of memory from IMRI binaries. We stress, however, that the agreement of the memory computation between models does not imply that the oscillatory waveforms themselves will necessarily agree, and building high-accuracy and extensive IMRI waveform models is an open and active area of work \cite{Wardell:2021fyy,Rosato:2021jsq,Nagar:2022icd}. It will be important to repeat this study once these models become available.

To assess the detectability of the memory signal in current and future gravitational wave detectors (primarily considering Advanced LIGO and ET), we compute both the optimal and angle-averaged signal to noise ratios (SNRs) for different binary configurations. Specifically, we have explored the SNR's dependence on the total mass, mass ratio, spin of the primary black hole, and eccentricity using memory signals with and without including subdominant harmonic modes. We find that memory signals become stronger when the primary black hole has positive spins, with the SNR growing by as much as a factor of $10$ as the spin is varied from $\chi \leq 0$ to $\chi \approx .99$. Fig.~\ref{Fig:memory_spin_snr} shows that memory signals from nonspinning BBH systems far from the detectability threshold may be detected for spins near $\chi \approx 0.95$. We find that when mild to moderate eccentricity is introduced, memory signals show rich structures (see Figures \ref{Fig:memory_diffecc} and \ref{Fig:memory_ecc}) -- with additional modulations both in the displacement and spin memory. However, the memory signal's amplitude hardly changes due to eccentricity, and consequently the SNRs for the memory modes in different eccentric configurations remain largely unchanged (see Table~\ref{tab:2}). We've also explored the SNR's dependence on mass ratio, largely confirming the overall expectation of that the memory signal (and hence SNR) will become weaker as the mass ratio increases. This follows from the fact that the oscillatory mode's amplitude decreases as $1/q$ in the large-mass-ratio limit. This trend is seen most clearly in Fig.~\ref{Fig:memory_diffq_snr} and continues almost up to equal-mass systems.

All of our main results have been obtained using the higher multipole displacement memory model~ \cite{Talbot:2018sgr} using an oscillatory waveform model with $(\ell,m)={(2,2),(2,1),(3,3),(3,2),(3,1),(4,2),(4,3)}$ modes in our computation. Unlike most previous works, our displacement memory results include contributions from higher-order modes that have been typically omitted in similar studies. We have therefore provided some comparisons to displacement memory effects computed with the (2,2) mode only. The inclusion of subdominant modes in the memory computation will ``activate" modes such as the (3,1) memory mode, which (for non-spinning systems) we see from Fig.~\ref{Fig:memory_modes_NRHybSur3dq8} has maximum power around $q \approx 2.5$. A representative sample of the mode hierarchy is shown in Figs.~\ref{Fig:memory_modes_q10} and \ref{Fig:memory_modes_q1075}, which shows the orbital-plane symmetry obeyed by the $(\ell,m)$ and $(\ell,-m)$ oscillatory modes is also obeyed by the memory modes. We find that including subdominant modes has a small but non-negligible impact on the systems considered here. Table \ref{tab:1} directly compares SNR values with and without higher order modes finding a difference by about $\sim 7\%$ across different detectors.

Our results indicate that displacement memory effects in IMRIs could be detected in future generation detectors such as the Einstein Telescope. Detection in current generation detectors would, however, require some amount of luck (e.g. systems merging very close and/or with a large, positive spin on the primary) and/or combining many events to compute the evidence (similar to Ref.~\cite{Lasky:2016knh}). Figure~\ref{Fig:q10_snrsummary}, for example, identifies regions of the total mass/distance parameter space where memory from a $q=10$ binary may have SNRs above 5 using a single advanced LIGO detector operating at design sensitivity. On the other hand, the spin memory would still be difficult to detect even for highly spinning, optimally oriented systems. Furthermore, as our SNR computations have been done assuming only one detector, repeating this study using a network of detectors would naively increase the memory SNRs by a factor of $\propto\sqrt{N_{\rm det}}$ where $N_{\rm det}$ is the number of GW detectors. However, a full study would be needed to include each detector's PSD as well as the relative orientation factors that may suppress or enhance any particular detector's sensitivity to the incoming memory signal. 

Heavy binaries with large positive spins are particularly promising for memory detections. For hierarchical mergers with second-generation (or higher) component black holes, spins of up to $\chi\approx.9$ are expected~\cite{doctor2020black,gerosa2021hierarchical}. The increased SNR at larger spin values would favorably contribute to forecasts for memory detection. However, to the best of our knowledge, all memory forecasts that have appeared in the literature (e.g.~\cite{Lasky:2016knh,Boersma:2020gxx}) use population models that favor nonspinning systems similar to GW190514. We expect that future work on memory forecasts that include both subdominant modes and mixed population models (1g+1g, 1g+2g, etc.) may find more optimistic forecasts. 

Finally, we note that in this work we have focused on IMRIs composed of a stellar origin BH and an IMBH.
It is also possible to form IMRIs with the combination of an IMBH and a massive BH.
Though their event rates are very uncertain, these are exciting and potentially very high SNR sources for the LISA detector \cite{AmaroSeoane:2007aw}. Their high SNR should also make them good candidates for detecting GW memory.


\begin{acknowledgments}
We thank Everett Gaige Field, Hamish Warburton, and David Nichols for helpful discussions and interactions throughout this work,
Colm Talbot for assistance with the Python package \texttt{GWMemory}, Keefe Mitman for assistance with the \texttt{sxs} Python package's memory module, and Alex Grant for assistance with a private Python package that we refer to in this paper as \texttt{GWForecasts}.
We further thank Colm, Keefe, and Alex for their help in facilitating a comparison documented in Appendix~\ref{app:code-comparison}.
We also thank an anonymous referee for raising important questions on an earlier version of our paper.
A portion of this work was carried out while a subset of the authors were in residence at the Institute for Computational and Experimental Research in Mathematics (ICERM) in Providence, RI, during the Advances in Computational Relativity program.  ICERM is supported by the National Science Foundation under Grant No. DMS-1439786.  
Simulations were performed on CARNiE at the Center for Scientific Computing and Visualization Research (CSCVR) of UMassD, which is supported by the ONR/DURIP Grant No.\ N00014181255 and the MIT Lincoln Labs {\em SuperCloud} GPU supercomputer supported by the Massachusetts Green High Performance Computing Center (MGHPCC). 
This research was supported in part by the Heising-Simons Foundation, the Simons Foundation, and National Science Foundation Grant No. NSF PHY-1748958.
The authors acknowledge support of NSF Grants No.~PHY-2106755 (G.K), No. PHY-1806665 (T.I.~and S.F), and No. DMS-1912716 (T.I., S.F, and G.K). 
T.I.~acknowledges additional support from the Kavli Institute for Theoretical Physics, University of California, Santa Barbara through Kavli Graduate Fellowship. 
N.W.~acknowledges support from a Royal Society–Science Foundation Ireland Research Fellowship. This publication has emanated from research conducted with the financial support of Science Foundation Ireland under Grant number 16/RS-URF/3428. 
\end{acknowledgments}  

\appendix
\section*{Appendix}
\label{appendix}
\section{Comparison of sxs, GWMemory, and GWForecasts}
\label{app:code-comparison}

Throughout this paper, we have performed numerical computations of the non-linear displacement memory from oscillatory gravitational waveform modes.  This direct computation, which is expected to be more accurate than analytic approximations, takes into account coupling between modes entering the energy flux expression~\eqref{eq:memory}.

Despite many appealing properties of a purely numerical memory computation, the associated software is sufficiently complicated that its helpful to have cross checks between independently written codes using different approaches. Throughout our paper, we have used the Python package $\texttt{GWMemory}$ \cite{gwmemory}, which is also a highly efficient code as it precomputes angular integrals appearing in Eq.~\eqref{eq:memory}. After the initial draft of our paper was completed, we became aware of two other codes that compute the displacement memory by a different means. One (currently non-public) code was recently used to compile forecasts (updating previous ones from Ref.~\cite{Boersma:2020gxx})for how long current and future detectors will need to operate in order to measure memory from binary black hole populations~\cite{grant2022outlook}. We will refer to this code as $\texttt{GWForecasts}$. A second code, a submodule of the $\texttt{sxs}$~\cite{sxs_code} Python package, implements techniques developed in the paper {\em Adding Gravitational Memory to Waveform Catalogs using BMS Balance Laws} by Mitman et al~\cite{mitman2021adding}.
The $\texttt{sxs}$'s memory submodule is designed to add memory to SXS waveform data obtained from the public catalog~\cite{Boyle:2019kee}. We wrote a wrapper around this submodule (implemented in $\texttt{gwtools}$ version 1.14) to allow users to generate memory when the waveform modes are instead represented as a Python dictionary. This allows the memory to be easily computed as \\[10pt] 
\begin{verbatim}
import numpy as np
from gwmemory import time_domain_memory as tdm
from gwtools import sxs_memory
import gwforecasts.memories.displacement as gwf
import gwsurrogate as gws

disp  = gwf.DisplacementMemory(ell_max=5)
model = gws.LoadSurrogate("NRHybSur3dq8")

chi = [0,0,0] # nonspinning BBH
t   = np.arange(-10000,100,.1)

# mass ratio 8 system without spin
t, h, dyn = model(8,chi,chi,times=t,f_low=0)

h_mem, times         = tdm(h_lm=h, times=t)
h_mem_forecasts      = disp.filter(h, t)
h_mem_sxs, times_sxs = sxs_memory(h, t)
\end{verbatim}
\vspace{10pt}

In this appendix, we briefly compare memory computed with $\texttt{GWMemory}$ (specifically, with a git commit hash of 2a4b8084144b13a3f542b1e132d9bc629d4ec9c1), $\texttt{sxs}$ (specifically version 2022.4.3), and $\texttt{GWForecasts}$.

\begin{figure}[t]
	\includegraphics[width=\columnwidth]{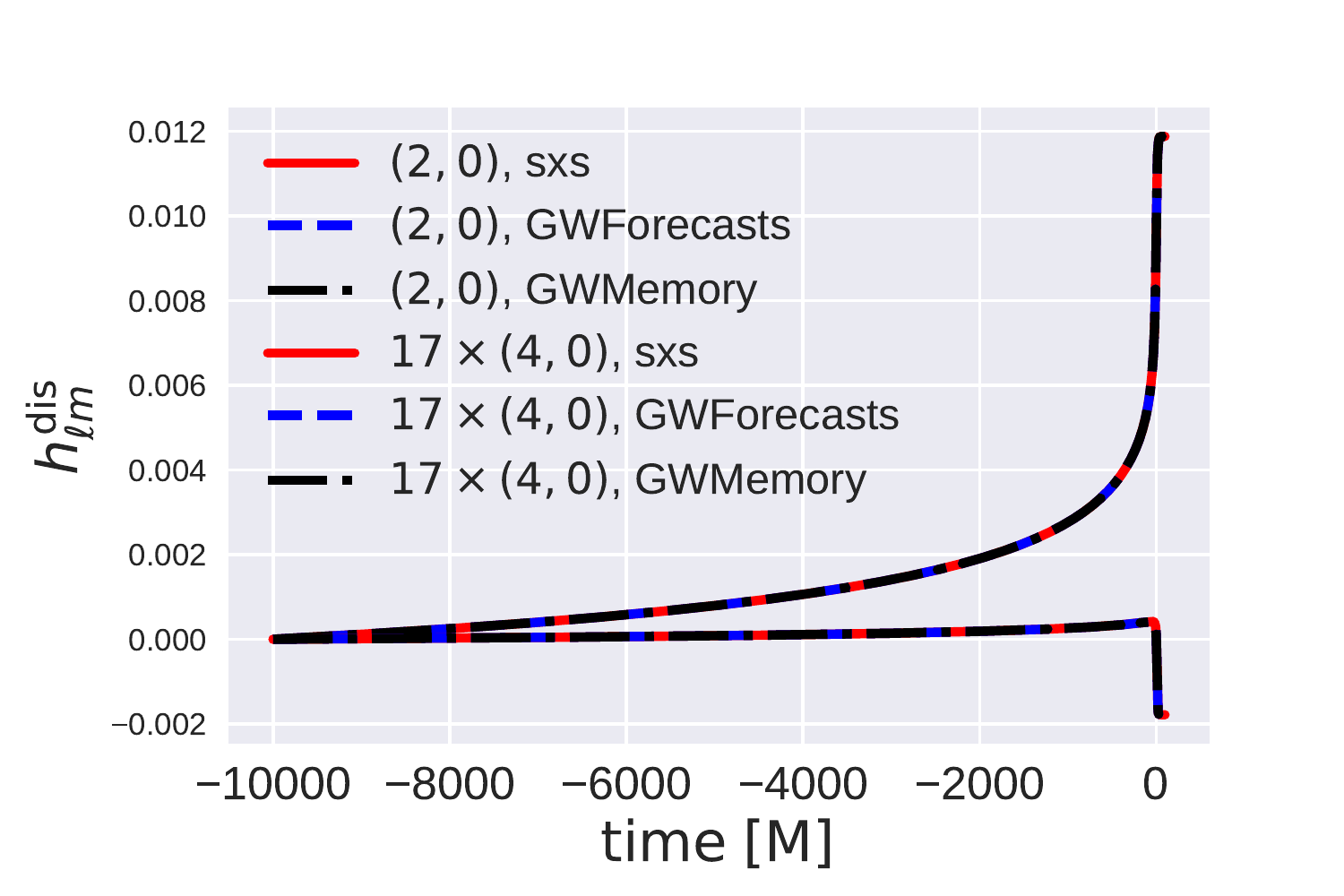}
	\caption{Displacement memory modes for a non-spinning BBH with mass ratio $q = 8$ computed using
		the $\texttt{sxs}$ (solid lines), $\texttt{GWForecasts}$ (dashed lines),
		and $\texttt{GWMemory}$ (dash-dot lines lines) packages. 
		All available $\ell \leq 5$
		oscillatory ``input" modes are passed to these packages, which in turn perform 
		a purely numerical computation of the memory. We find broad agrement between
		all three codes. Here we show the dominant $(2,0)$ 
		as well as the $(4,0)$ mode, where we have multiplied the mode's amplitude 
		by $17$ for visual assistance. Subdominant modes (not shown) are also in agreement.
		\label{Fig:app}
	}
\end{figure}

Fig.~\ref{Fig:app} 
compares the $\texttt{sxs}$ (solid red lines), $\texttt{GWMemory}$ (dashed-dot black lines),
and $\texttt{GWForecasts}$ (dashed blue lines)
computation for the displacement memory's $(2,0)$ and $(4,0)$ modes. 
The input waveform comes from the \texttt{NRHybSur3dq8} model 
for a $q=8$, nonspinning BBH system.
Both the $(2,0)$ and $(4,0)$ modes are visually identical among all three codes, as 
are many of the subdominant modes (not shown). In earlier versions of $\texttt{GWMemory}$ (used to compile results from our paper's arXiv version 1) we noticed small differences in the $(2,0)$ mode and sometimes large discrepancies in the subdominant modes.


\section*{References}
\bibliography{EMRIMem}

\end{document}